\documentclass[12pt,groupedaddress,pra,aps]{revtex4-1}
\usepackage[utf8]{inputenc}
\usepackage{amsmath}
\usepackage{amsfonts}
\usepackage{amssymb}
\usepackage{amsthm}
\usepackage{color,graphicx}
\usepackage{graphics}
\usepackage{subfigure}
\usepackage{color}
\usepackage{epsfig}
\usepackage{bm}

\DeclareMathOperator\erf{erf}
\DeclareMathOperator\sinc{sinc}

\begin{document}

\author{Marko Toro\v{s}}
\email{marko.toros@ts.infn.it}
\author{Angelo Bassi}
\email{bassi@ts.infn.it}

\affiliation{Department of Physics, University of Trieste, 34151 Miramare-Trieste, Italy \\
Istituto Nazionale di Fisica Nucleare, Sezione di Trieste, Via Valerio 2, 34127 Trieste, Italy}

\title{Bounds on Collapse Models from Matter-Wave Interferometry: Calculational details}

\date{\today}

\begin{abstract}
We present a simple derivation of the interference pattern in matter-wave interferometry as predicted by a class of master equations, by using the density matrix formalism. We apply the obtained formulae to the most relevant collapse models, namely the Ghirardi-Rimini-Weber (GRW) model, the continuous spontaneous localization (CSL) model together with its dissipative (dCSL) and non-markovian generalizations (cCSL), the quantum mechanics with universal position localization (QMUPL) and the Di\'{o}si-Penrose (DP) model. We discuss the separability of the collapse models dynamics along the 3 spatial directions, the validity of the paraxial approximation and the amplification mechanism.
We obtain analytical expressions both in the far field and near field limits. These results agree with those already derived in the Wigner function formalism. 

We compare the theoretical predictions with the experimental data from two relevant matter-wave experiments: the 2012 far-field experiment and the 2013 Kapitza Dirac Talbot Lau (KDTL) near-field experiment of Arndt's group. We show the region of the parameter space for each collapse model, which is excluded by these experiments. We show that matter-wave experiments provide model insensitive bounds, valid for a wide family of dissipative and non-markovian generalizations. 
\end{abstract}

\maketitle 

\section{Introduction}\label{introduction}

The interest in exploring, not only theoretically but also experimentally, the foundations of quantum mechanics has significantly increased over the years. After the establishment of quantum nonlocality, first with the famous works of J. Bell \cite{bell1964einstein,bell2004speakable} and subsequently with the experimental confirmation done by the group of A. Aspect \cite{PhysRevLett.47.460,PhysRevLett.49.91,PhysRevLett.49.1804}, perhaps the most relevant question is if the collapse of the wave function is a physical phenomenon or not. Quantum Mechanics predicts that also macroscopic objects---being made of atoms, which are quantum---should live in the superpositions of different states. But this has never been observed. Why so? Is it simply because macroscopic quantum superpositions are difficult to spot, due to environmental noises, or because they are forbidden for some physical reason? No one knows the answer yet, and research is active in testing which of the two alternatives is correct.

Collapse models \cite{Bassi2003257,RevModPhys.85.471} have been formulated to take this second possibility into account: nature forbids macroscopic systems to live in superposition states. From the mathematical point of view, the Schr\"odinger equation is modified by adding nonlinear and stochastic terms, which account for the quantum-to-classical transition. For microscopic systems, the standard  quantum evolution is the dominant contribution to the dynamics, hence they behave in a fully quantum way, as repeatedly confirmed in experiments. For macroscopic objects, on the other hand, the opposite is true: the nonlinear terms prevent superpositions to occur. The border between these two regimes lies somewhere in the mesoscopic world.

As such, collapse models are predictively different from standard quantum mechanics, and research is active in testing them \cite{RevModPhys.85.471}, because any test of collapse models is a test of the quantum superposition principle, which lies at the foundations of any quantum theory. 

Different collapse models have been proposed over the years. The most famous model is the Continuous Spontaneous Localisation (CSL) model \cite{PhysRevA.39.2277,PhysRevA.42.78}, a generalisation of the original Ghirardi-Rimini-Weber (GRW) model \cite{PhysRevD.34.470} to systems containing identical particles. The CSL model, in the limit of short superposition distances, reduces to the Quantum-Mechanics-with-Universal-Position-Localization (QMUPL) model \cite{PhysRevA.40.1165,PhysRevA.42.5086}. In all cases, the noise driving the collapse is a white noise. This modelling of the noise is very useful from the practical point of view, as in this case the equations of motions are relatively simple, however a white noise is not physical. For this reason, in recent years the CSL model has been generalised in two directions. On the one side, dissipative effects have been included in the dynamics, which drive any quantum system, during the collapse, to a thermal state. This partly solves the problem of the steady energy increases, which affects the CSL model. The model is called the dissipative CSL (dCSL) model \cite{Smirne:2014paa}. Its limiting case, the dissipative QMUPL model, has also been studied \cite{0305-4470-38-37-007}. On the other hand, the white noise has been replaced by a coloured noise \cite{PhysRevA.48.913,1751-8121-40-50-012, 1751-8121-41-39-395308}. In this case we speak of coloured CSL (cCSL). A coloured noise introduces non-Markovian terms in the dynamics, making the whole mathematical analysis rather difficult. The cCSL model reduces to the coloured QMUPL model in the limit of short superposition distances \cite{PhysRevA.80.012116}. Only for  the QMUPL model, both dissipative and non-Markovian effects have been combined together in a single model \cite{PhysRevA.86.022108}, so far. Independent from the CSL model, there is the Diosi-Penrose (DP) model \cite{PhysRevA.40.1165}, which is a first attempt to link the collapse of the wave function to gravity. 

All these models contain phenomenological parameters. The GRW and CSL models are defined in terms of a localization rate $\lambda$ and a localization length $r_C$. $\lambda$ gives the frequency of the localization events for a reference object of mass $m_0=1$ amu, while $r_C$ describes how well an object is localized. The QMUPL model has only the parameter $\eta$, which can be related to the GRW/CSL parameters~\cite{durr2011stochastic}. A common open question of both the GRW and CSL models, is to explain the origin of the noise in the dynamical equations. A first attempt at addressing this issue is given by the DP model, where the strength of the localization is set by the  gravitational interaction through the gravitational constant $G$. The DP model introduces only one cut-off length phenomenological parameter $R_0$, which cures the ultraviolet divergence of the gravitational interaction. The effective collapse rate, analogous to $\lambda$, is given by $G m_0^2/\sqrt{\pi} \hbar  R_0$, while $R_0$ describes how well an object is localized, analogous to $r_C$. 

One unwanted feature, common to the GRW, CSL and DP  models, is the energy divergence for very long (cosmological) times. Indeed, the GRW, CSL and DP master equations have the structure of a  quantum linear Boltzmann equation \cite{Vacchini200971} of a particle immersed in a infinite temperature bath. One attempt to solve this issue, is proposed by the dissipative extensions of the GRW and CSL models, namely the dGRW and dCSL models, respectively \cite{PhysRevA.90.062135,Smirne:2014paa}. Here the energy divergence is eliminated by the introduction of a noise temperature parameter $T$. Another approach to solve the energy divergence, adopted by the cCSL model \cite{1751-8121-40-50-012,1751-8121-41-39-395308}, is to replace the non-physical white noise by a colored noise with a finite correlation time parameter $\tau_C$. We provide a brief summary of these models in section \ref{socm}. 

Bounds on collapse models parameters are first investigated in~\cite{1751-8121-40-12-S03} and an overview is given
in~\cite{adler2009quantum, 1751-8121-45-6-065304}. In this paper, we complete and improve the previous analysis of collapse models' predictions for matter-wave interferometry. The bounds on the parameters $(\lambda,r_C)$ can be conveniently studied in the parameter space \cite{collett1, collett2} shown in Fig.~\ref{CSL_parameter_diagram}, while the bounds on the parameters $\eta$ are shown in Figs.~\ref{fig_QMUPL}. We obtain bounds for all the collapse models introduced above, from the localization requirement of macroscopic objects and from experimental data.

We describe how to obtain the bounds from the localization requirement of macroscopic objects at the end of section \ref{sec_amp}, where we discuss a key feature of all collapse models, namely  the amplification of the effective collapse rate, as the size (mass) of the system increases. In other words, under standard assumptions, the center of mass motion for a rigid many-body system is governed by the single particle equation with a rescaled collapse rate. 

We show that, on the one hand, current matter-wave experiments do not give significant bounds on the DP parameter $R_0$, while, on the other hand, the localization requirement of macroscopic objects (as defined in Secs.~\ref{sec_amp} and \ref{desec}), excludes all values of $R_0$.

The first and main prediction of collapse models is the gradual modification of the interference pattern in interferometric experiments, as the mass of the diffracted object becomes large. Therefore matter-wave experiments provide the most direct test of collapse models. They are described in sections \ref{dcmip_qm}, \ref{desec}. 
\begin{figure}[!htb]
\begin{center}
\includegraphics[width=1.0\textwidth,trim={50 150 100 75},clip]{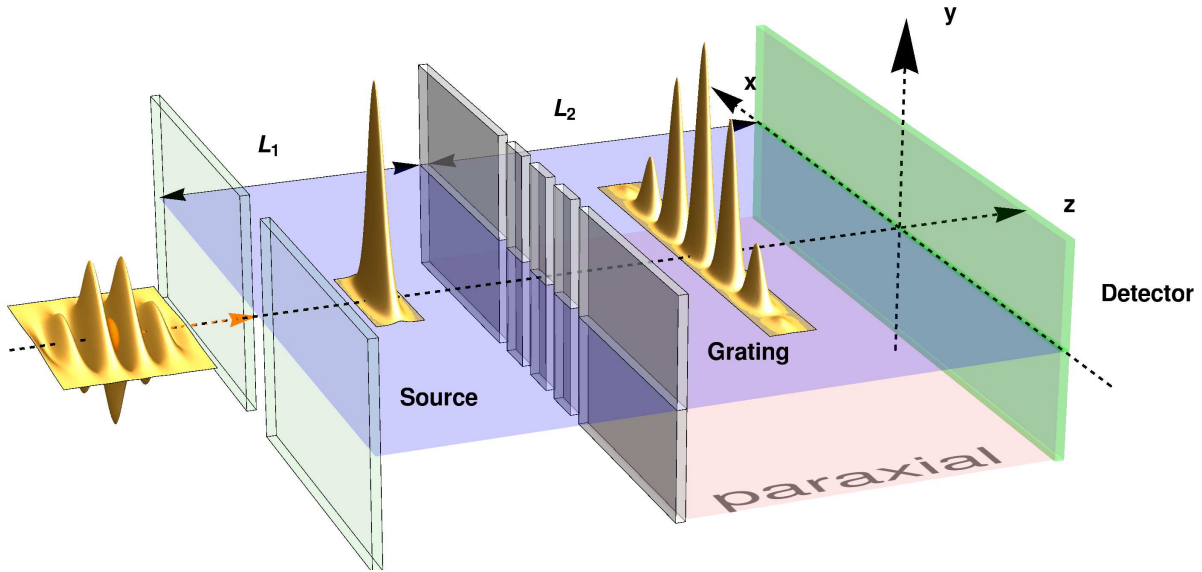}
\caption{The common structure of far-field  and near-field diffraction experiments. A molecular beam from an incoherent source  propagates along the $z$ axis. Each molecule is emitted from the source, propagates to the grating, where it is diffracted and then recorded by the detector. The molecules individually recorded gradually form an interference pattern. The figure shows a mechanical grating (with $N=4$ slits), but the analysis in this paper applies to more general gratings, e.g. optical gratings. The distance from the source to the grating is $L_1$ and the distance from the grating to detector is $L_2$. Between the grating and the detector we identify the paraxial (Fresnel) regime.}
\label{fig_paraaxial}
\end{center}
\end{figure}
In particular, in section \ref{dcmip_qm} we derive the interference pattern $p(x)$ for the experimental setup described in Fig.~\ref{fig_paraaxial}. We compare the theoretical interference patterns with experimental data from the 2012 far-field  matter-wave interferometry experiment \cite{MSCH} and the 2013 Kapitza Dirac Talbot Lau (KDTL) near-field  matter-wave interferometry experiment \cite{C3CP51500A}, both performed by Arndt's group in Vienna, in section \ref{desec}. Tests of the CSL model with matter-wave interferometry experiments were first investigated in \cite{1751-8121-40-12-S03,StefanNimmrichter}, in particular, in the context of the OTIMA experiment \cite{PhysRevA.83.043621}.

The interference pattern derived in section \ref{dcmip_qm} in the density matrix formalism, is not limited only to collapse models, but is valid for a large class of dynamics. In particular, we also discuss under which conditions the diffraction experiment can be reduced to a one dimensional problem, since the collapse dynamics, unlike ordinary quantum mechanics, is not separable in the three spatial dimension, even for the free particle dynamics. In this way we justify the calculation of the interference pattern in the paraxial approximation. In addition, the density matrix formalism outlines the similarities of far-field and near-field interference, by presenting a unified derivation. We also reobtain the results for diffraction experiments that were derived in the Wigner function formalism \cite{PhysRevA.83.043621,PhysRevA.70.053608, PhysRevA.73.052102}. 

In section \ref{conclusions} we combine all the parameter bounds for the CSL, GRW, dCSL, dGRW and cCSL models in a single parameter diagram and we discuss the bounds on the parameter of the QMUPL model as well as on the length parameter of the DP model.

\clearpage
\section{Derivation of the interference pattern}\label{dcmip_qm}

For all collapse models considered here (see the next Section), the evolution of the free single-particle density matrix has the form:
\begin{equation}\label{rho_3d}
\rho(\bm{x},\bm{x'},t)=\frac{1}{(2\pi\hbar)^3} \int d\bm{\tilde{k}}  \int \bm{\tilde{w}} e^{-\frac{i}{\hbar}\bm{\tilde{k}} \cdot \bm{\tilde{w}}}\\
 F(\bm{\tilde{k}},\bm{x}-\bm{x'},t) \rho^{\text{\tiny QM}}(\bm{x}+\bm{\tilde{w}},\bm{x'}+\bm{\tilde{w}},t),
\end{equation}
where $\rho^{\text{\tiny QM}}$ is the free standard quantum mechanical density matrix and the function $F$ depends on the type of collapse model. 

The quantum mechanical description of matter-wave interferometry is usually treated as a one-dimensional problem. This is justified by the fact that the free Schr\"odinger dynamics is separable along the three directions of motion. On the contrary, in general the  dynamics given by Eq.~\eqref{rho_3d} is not separable, not even in the free particle case. We show, however, that due to the specific geometry and experimental parameters of the diffraction experiments here considered, we can effectively separate the collapse dynamics in the three spatial directions, thus considerably simplifying the problem. Along with this, we will investigate the assumptions that are required for the justification of the one dimensional approximation. Actually, it is instructive to first carry out the calculation in the 1D (paraxial) approximation before justifying it. 

The derivation of the paraxial interference  pattern is the main result of this section. We then apply the paraxial interference formula to the far-field and near-field experimental setups. In order to simplify the comparison with similar results obtained in the literature, we will adopt the notation of \cite{1464-4266-5-2-362}. We will also omit the overall normalization factors for the wave functions, density matrices and probability densities. At any step of the calculation, one can obtain a normalized quantity by dividing with an appropriate normalization factor. \\ \\

\noindent {\bf Paraxial approximation}.
We first review the quantum mechanical derivation of the interference pattern in the paraxial (Fresnel) region, as depicted in Fig.~\ref{fig_paraaxial}. We label with $z_1, z_2, z_3$ the positions of source, grating and detector along the optical axis $z$, respectively. Similarly, we label the horizontal coordinates along the optical elements as $x_1$, $x_2$, $x_3$, respectively.  

In the paraxial diffraction region the evolution of the wave function can be approximated by the free quantum mechanical wave function propagation in one spatial dimension~\footnote{This coincides with the Fresnel diffraction integral.}:
\begin{equation}\label{psi_qm}
\psi(x;t=L/v)=\int_{-\infty}^{+\infty} dx_0 \psi_0(x_0) e^{\frac{ik}{2L}(x-x_0)^2},
\end{equation}
where $k$ is the wave number of the matter wave, $\psi_0$ is the initial wave function and $\psi$ is the wave function after it has propagated for a distance $L$ in a time $t=L/v$, where $v$ is the speed of propagation along the optical axis $z$. One has the usual relation $mv=\hbar k$, where $m$ is the mass of the system (the macromolecule). 
In the language of density matrices Eq.~\eqref{psi_qm} reads:
\begin{equation}\label{rho_qm}
\rho^{\text{\tiny QM}}(x,x';t=L/v)=\int_{-\infty}^{+\infty}dx_0' \int_{-\infty}^{+\infty} dx_0 \rho_0(x_0,x_0')
e^{\frac{ik}{2L}((x-x_0)^2-(x'-x_0')^2)},
\end{equation}
where $\rho_0(x_0,x_0')$ is the initial density matrix and $\rho^{\text{\tiny QM}}(x,x';t)$ is the density matrix after it has propagated for a distance $L$ in a time $t=L/v$.

The calculation of the interference pattern can be summarized in the following steps. 
\begin{description}
\item[$\boldsymbol{[z_1]}$] We choose the initial wave function at $z_1$. Both the far-field and near-field experiments will be modeled by a completely incoherent source at $z_1$, meaning that the wave functions associated to different molecules are uncorrelated and spatially localized initially. It is then sufficient to consider a single source at point $(x_1, z_1)$. At the end, one can integrate over the extension of the source. The corresponding initial wave function is given by
\begin{equation}\label{eq1qm}
\psi_1(\tilde{x}_1)=\delta(x_1-\tilde{x}_1).
\end{equation}
\item[ $\boldsymbol{[z_1}$ to $\boldsymbol{z_2]}$] We propagate the wave function to $z_2$ according to Eq.~\eqref{psi_qm}:
\begin{equation}\label{eq2qm}
\psi_2(x_2)=\int_{-\infty}^{+\infty} d\tilde{x}_1 \psi_1(\tilde{x}_1) e^{\frac{ik}{2L_1}(x_2-\tilde{x}_1)^2}.
\end{equation}
\item[$\boldsymbol{[z_2]}$] We now assume that the optical element at position $z_2$ has a transmission function $t(x)$. The wave function immediately after the grating at $z_2$ is given by $ t(x_2)\psi_2(x_2)$.
\item[$\boldsymbol{[z_2}$ to $\boldsymbol{z_3]}$] We propagate the wave function from $z_2$ to $z_3$ according to Eq.~\eqref{psi_qm}: 
\begin{equation}\label{eq3qm}
\psi_3(x_3)=\int_{-\infty}^{+\infty} dx_2 t(x_2)\psi_2(x_2) e^{\frac{ik}{2L_2}(x_3-x_2)^2}.
\end{equation}
\item[$\boldsymbol{[z_3]}$] The detector records the arrival of the molecules along the axis $x_3$. The probability distribution is $p_3(x_3)=|\psi_3(x_3)|^2$. After combining the equations of the previous steps we obtain the interference pattern:
\begin{equation}\label{p3_qm}
\begin{split}
 p_3(x_3)= &\int_{-\infty}^{+\infty} dx_2 \int_{-\infty}^{+\infty}  dx'_2 t(x_2)t^{*}(x'_2)\\
& e^{-\frac{ik}{2L_2}(x_2-x'_2)x_3} 
 e^{\frac{ik}{2L_1}(x_2^2-{x'_2}^2)}e^{\frac{ik}{2L_2}({x'_2}^2-x_2^2)} e^{-\frac{ik}{L_1}(x_2-x_2')x_1}.
\end{split}
\end{equation}
\end{description}
Note that Eq.~\eqref{p3_qm} was derived from Eq.~\eqref{psi_qm}, but it could equally well be derived from the density matrix evolution given by Eq.~\eqref{rho_qm}. 

We now consider what happens if in place of the standard quantum evolution, we have the following density matrix evolution:
\begin{equation}\label{1d_rho}
\rho(x,x';t)=\frac{1}{2\pi\hbar} \int_{-\infty}^{+\infty} d\tilde{k}  \int_{-\infty}^{+\infty} \tilde{w} e^{-\frac{i}{\hbar}\tilde{k} \tilde{w}}
F(\tilde{k},0,0;x-x',0,0;t) \rho^{\text{\tiny QM}}(x+\tilde{w},x'+\tilde{w};t).
\end{equation}
We will justify Eq.~\eqref{1d_rho} below, when we discuss the separability issue. The calculation of the interference pattern can be again carried out as before.
\begin{description}
\item[$\boldsymbol{[z_1]}$] We  consider a single source at point $(x_1, z_1)$. The corresponding initial wave function is given by $\psi_1(\tilde{x}_1)=\delta(x_1-\tilde{x}_1)$ and the corresponding density matrix is given by 
\begin{equation}\label{grwdp1}
\rho_1(\tilde{x}_1,\tilde{x}'_1)=\delta(x_1-\tilde{x}_1)\delta(x_1-\tilde{x}'_1).
\end{equation}

\item[$\boldsymbol{[z_1}$ to $\boldsymbol{z_2]}$] We propagate the density matrix from the point $z_1$ to the point $z_2$ along the optical axis using Eq.~\eqref{1d_rho}:
\begin{equation}\label{grwdp2}
\rho_2(x_2,x_2)=\frac{1}{2\pi\hbar} \int_{-\infty}^{+\infty} d\tilde{k}  \int_{-\infty}^{+\infty} \tilde{w} e^{-\frac{i}{\hbar}\tilde{k} \tilde{w}} 
 F(\tilde{k},0,0;x-x',0,0;t) \rho_2^{\text{\tiny QM}}(x_2+\tilde{w},x_2'+\tilde{w}),
\end{equation}
where according to Eq.~\eqref{rho_qm} and Eq.~\eqref{grwdp1}:
\begin{equation}
\rho_2^{\text{\tiny QM}}(x_2,x_2')=e^{\frac{ik}{2L_1}(x_2^2-x_2'^2)}e^{-\frac{ik}{L_1}(x_2-x_2')x_1}.
\end{equation}
In Eq.~\eqref{grwdp2} the $\tilde{w}$ integration yields a Dirac delta function  $\delta(\tilde{k}-\frac{\hbar k}{L_1}(x_2-x_2'))$ and hence after the $\tilde{k}$ integration we obtain:
\begin{equation}
\rho_2(x_2,x_2')= e^{\frac{ik}{2L_1}(x_2^2-x_2'^2)}e^{-\frac{ik}{L_1}(x_2-x_2')x_1}F\left(\frac{\hbar k}{L_1}(x_2-x_2'),0,0;x_2-x_2',0,0;t_1\right).
\end{equation}
\item[$\boldsymbol{[z_2]}$] We  apply the grating's transmission function $t(x)$ to the density matrix and obtain $t(x_2) \rho_2(x_2,x_2') t^*(x_2')$.
\item[$\boldsymbol{[z_2}$ to $\boldsymbol{z_3]}$] 
We  perform a free propagation according to Eq.~\eqref{1d_rho} from $z_2$ to $z_3$:
\begin{equation}\label{grwdp3}
\rho_3(x_3,x_3')=\frac{1}{2\pi\hbar} \int_{-\infty}^{+\infty} d\tilde{k}  \int_{-\infty}^{+\infty} \tilde{w} e^{-\frac{i}{\hbar}\tilde{k} \tilde{w}}  F(\tilde{k},0,0;x_3-x'_3,0,0;t_2) \rho_3^{\text{\tiny QM}}(x_3+\tilde{w},x_3'+\tilde{w})
\end{equation}
where 
\begin{equation}
\rho_3^{\text{\tiny QM}}(x_3,x_3')= \int_{-\infty}^{+\infty} dx_2\int_{-\infty}^{+\infty}  t(x_2) t^*(x_2') dx_2' \rho_2(x_2,x_2') e^{\frac{ik}{2L_2}((x_3-x_2)^2-(x_3'-x_2')^2)}.
\end{equation}
\item[$\boldsymbol{[z_3]}$] 
The interference pattern is again proportional to the probability density $p(x)=\rho_3(x,x)$. The $\tilde{w}$ integration yields a Dirac delta function  $\delta(\tilde{k}+\frac{\hbar k}{L_2}(x_2-x_2'))$. Hence after the $\tilde{k}$ integration we obtain the interference pattern:
\begin{equation}\label{p3_collapse}
\begin{split}
&p(x)=\int_{-\infty}^{+\infty} dx_2\int_{-\infty}^{+\infty}  dx_2' 
\; D(x_2-x_2') \;t(x_2)t^*(x_2') \\
&\times  e^{-i \frac{m v}{\hbar} (x_2- x_2')( \frac{x_1}{L_1}+\frac{x}{L_2}) } 
e^{i \frac{m v}{\hbar}\frac{L_1+L_2}{2L_1 L_2}(x_2^2-x_2'^2)},
\end{split}
\end{equation}
where 
\begin{equation}\label{Dfunc_sep}
D(x_2-x_2')=F(-\frac{\hbar k}{L_2}(x_2-x_2'),0,0;0,0,0;t_2)
F(\frac{\hbar k}{L_1}(x_2-x_2'),0,0;(x_2-x_2'),0,0;t_1).
\end{equation}
\end{description}
As we can see, the interference pattern in Eq.~\eqref{p3_collapse} differs from the pure quantum mechanical interference pattern of Eq.~\eqref{p3_qm} by the presence of $D(x_2-x_2')$. \\ \\

\noindent {\bf Separability}. We now perform the full 3D treatment of the problem to justify the 1D approximation. We consider an initial Gaussian wave packet, evolving according to the full dynamics in Eq.~\eqref{rho_3d}. For the geometry, we refer again to the experimental setup depicted in Fig.~\ref{fig_paraaxial}. 
\begin{figure}[!htb]
\begin{center}
\includegraphics[width=0.25\textwidth,trim={0 0 0 0},clip]{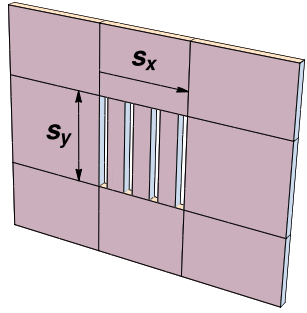}
\caption{The grating has non-zero transmission function limited to a rectangle of size $s_x \times s_y$, e.g. here we show a mechanical grating with $N=4$ slits with total horizontal extension $s_x$ and slit height $s_y$. The analysis of this section applies also to other types of gratings, e.g. an optical grating.}
\label{fig_paraaxial_supp}
\end{center}
\end{figure}
We will show under which assumptions the interference pattern is given by Eq.~\eqref{p3_collapse}, thus justifying the above analysis in the 1D (paraxial) approximation. The assumptions are:
\begin{description}
\item[1]
The extension of the macromolecule $\sigma(t)$ is much smaller then the distances $L_1$, $L_2$ during the time of flight $t$: 
\begin{equation}
\sigma(t) \ll L_1,L_2.
\end{equation}
This key assumption allows to split the flight of the molecule from the source at time $t=0$ to the grating at time $t_1$, from the motion from the grating at time $t_1$ to the detector at time $t_1+t_2$, and to treat the non-free interaction with the grating as  instantaneous. This is necessary in order to conveniently  introduce a transmission function for the grating $t_{xy}(x,y)$. In particular, we choose $t_{xy}(x,y)=t(x)t_y(y)$, where
\begin{equation}\label{txx}
t(x)=0 \;\;\text{if $|x| > \frac{s_x}{2}$},
\end{equation}
while for $|x| < \frac{s_x}{2}$ it depends on the type of grating and
\begin{equation}\label{tyy}
t_y(y)=\begin{cases}
    1, & \text{if $|y| \leq \frac{s_y}{2}$}.\\
    0, & \text{if $|y| > \frac{s_y}{2}$}.
  \end{cases}
\end{equation}
where $s_x$, $s_y$ are described in Fig.~\ref{fig_paraaxial_supp} and $t(x)$ is to be identified with the transmission function used above, when working in the 1D (paraxial) approximation.

\item[2]
We assume that the molecule extension $\sigma_2$ at time $t_1$, as it reaches the grating, is much larger than the molecule extension $\sigma_1$, at time $t=0$, as it leaves the source:
\begin{equation}\label{2nd2}
\sigma_1 \ll \sigma_2.
\end{equation}
\item[3]
We require that the grating transmission function satisfies (see Fig.~\ref{fig_paraaxial_supp}):
\begin{align}
s_x \ll \sigma_2, \label{sxassumption1}\\
\sigma_2 \ll s_y.\label{syassumption1}
\end{align}
\end{description}

Using ordinary quantum mechanics it is easy to give an estimate for the molecule extension at the grating: $\sigma_2=\frac{\hbar t_1}{m \sigma_1}$ with $\sigma_1$ the extension at the source (see analysis below). Using this relation let us check the validity of the above assumptions for the two experiments considered. 

For the far-field diffraction experiment \cite{MSCH} we have $L_1 = 0.702\text{m}$, $L_2=0.564\text{m}$, $s_x=3 \mu\text{m}$, $s_y=60\mu\text{m}$  and  the molecular speed along the $z$ axis $v \sim 100\text{ms}^{-1}$. The above assumptions are satisfied if the initial molecular extension at the source is contained in the interval $4 \times 10^{-9}\text{m}$ $\lesssim  \sigma_1 \lesssim  7\times 10^{-8}\text{m}$. No one knows the actual value of $\sigma_1$. The range of values here considered makes the initial spread much smaller than the extension of the source ($s=1\mu\text{m}$) as given by the collimator and also provides a justification as to why the source is incoherent. 

For the near-field KDTL diffraction experiment \cite{C3CP51500A} we have $L_1=L_2=10.5 \text{cm}$, while it is difficult to give estimates for parameters $s_x$, $s_y$ of the light grating. Anyhow, making the following guess for these  parameters: $s_x= 10^{-3}\text{m}$, $s_y= 100 \times 10^{-3}\text{m}$ (and being the molecular speed along the z axis $v \sim 100\text{ms}^{-1}$), the above assumptions are satisfied if the initial molecular extension at the source is contained in the interval $10^{-13}\text{m} \lesssim  \sigma_1 \lesssim  10^{-11}\text{m}$. This is to be compared with the slit openings of the source grating $l=110nm$. Without more precise estimates for the molecule extension $\sigma(t)$ it is difficult to assess the validity of the above assumptions and hence of the 1D approximation. We stress  that we are considering a single molecule emitted from the source. In particular, the single molecule extension $\sigma(t)$ should not be confused with the spatial coherence length of the beam, which is a property of an ensemble of particles emitted from the source.

As in the previous section, the calculation of the interference pattern can be split into several steps.
\begin{description}
\item[$\boldsymbol{[z_1]}$] It is convenient to work in a boosted reference frame along the $z$ axis with molecular velocity $v$ \footnote{For non boost-invariant dynamics, one has to choose the correct function $F$ depending on the reference frame.}, i.e. moving alongside the molecule. To simplify the analysis we neglect gravity and we consider an initial Gaussian wave-function centered at $(x_1, 0,0)$: 
\begin{equation}\label{init_3d_psi}
\psi_1(\tilde{x}_1,\tilde{y}_1,\tilde{z}_1)=e^{-\frac{(x_1-\tilde{x}_1)^2}{4 \sigma_1^2}}
e^{-\frac{\tilde{y}_1^2}{4 \sigma_1^2}}
e^{-\frac{\tilde{z}_1^2}{4 \sigma_1^2}},
\end{equation}
with the corresponding density matrix given by 
\begin{equation}\label{init_3d_rho}
\rho_1(\tilde{x}_1,\tilde{y}_1,\tilde{z}_1 ;\tilde{x}'_1, \tilde{y}'_1,\tilde{z}'_1)=\psi_1(\tilde{x}_1,\tilde{y}_1,\tilde{z}_1)\psi_1^*(\tilde{x}'_1,\tilde{y}'_1,\tilde{z}'_1).
\end{equation}
\item[$\boldsymbol{[z_1}$ to $\boldsymbol{z_2]}$] We propagate the density matrix $\rho_1$ from $t=0$ to  $t=t_1=L_1/v$ using Eq.~\eqref{rho_3d}. We denote the resulting density matrix as $\rho_2(x_2, y_2, z_2;x'_2, y'_2, z'_2)$ (and by $\rho^{\text{\tiny QM}}_2(x_2, y_2, z_2;x'_2, y'_2, z'_2)$ the quantum mechanical evolution of $\rho_1$ at $t=t_1$). In particular, using the separability of ordinary quantum mechanics, the quantum mechanical wave function just before $t_1$ is given by:
\begin{equation}\label{grating_3d_2}
\psi_2^{\text{\tiny QM}}(x_2, y_2, z_2) = \psi_2^{\text{\tiny QM}(1)}(x_2;x_1) \psi_2^{\text{\tiny QM}(1)}(y_2;0) \psi_2^{\text{\tiny QM}(1)}(z_2;0),
\end{equation}
where $\psi_2^{\text{\tiny QM}(1)}(x_2;x_1)=\exp \left[-\frac{(x_2-x_1)^2}{4 \sigma_1^2 (1+\frac{i\hbar t_1}{2 m \sigma_1^2})} \right]$. Hence the quantum mechanical density matrix is given by:
\begin{equation}\label{grating_3d_rho}
\rho^{\text{\tiny QM}}_2(x_2, y_2, z_2;x'_2, y'_2, z'_2;t_1) = 
\rho^{\text{\tiny QM(1)}}_2(x_2,x'_2;x_1)
\rho^{\text{\tiny QM(1)}}_2(y_2,y'_2;0)
\rho^{\text{\tiny QM(1)}}_2(z_2,z'_2;0),
\end{equation}
where 
\begin{equation}\label{rho_3d_bg1}
\begin{split}
\rho^{\text{\tiny QM(1)}}_2(x,x',x_1)= \exp \left[\right.&
-\frac{1}{\sigma_2^2 } \left( \right. x^2 (1-\frac{i\hbar t_1}{2 m \sigma_1^2})
+ x'^2 (1+\frac{i\hbar t_1}{2 m \sigma_1^2}) \\
&-2 x x_1 (1+ \frac{i \hbar t_1}{2 m \sigma_1^2}) 
-2 x' x_1 (1 - \frac{i \hbar t_1}{2 m \sigma_1^2})    
+2 x_1^2 \left.\right)
\left.\right].
\end{split}
\end{equation}
and $\sigma_2=\frac{\hbar t_1}{m \sigma_1}$ because of Eq.~\eqref{2nd2}, i.e.
$\frac{\hbar^2 t_1^2}{4 m^2 \sigma_1^4} \gg 1$.
To summarize this step of the calculation:
\begin{equation}\label{rho_3d_bg}
\begin{split}
\rho_2(x_2, y_2, z_2;x'_2, y'_2, z'_2) = 
& \int d\tilde{k}_x \int d\tilde{w}_x \rho^{\text{\tiny QM(1)}}_2(x_2 + \tilde{w}_x ,x'_2 +\tilde{w}_x;x_1) e^{-\frac{i}{\hbar}\tilde{k}_x \tilde{w}_x}  \\
&\times\int d\tilde{k}_y \int d\tilde{w}_z \rho^{\text{\tiny QM(1)}}_2(y_2 +\tilde{w}_y ,y'_2 +\tilde{w}_y;0) e^{-\frac{i}{\hbar}\tilde{k}_y \tilde{w}_y} \\
&\times \int d\tilde{k}_z \int d\tilde{w}_z \rho^{\text{\tiny QM(1)}}_2(z_2+\tilde{w}_z,z'_2+\tilde{w}_z;0) e^{-\frac{i}{\hbar}\tilde{k}_z \tilde{w}_z} \\
&\times F(\tilde{k}_x,\tilde{k}_y,\tilde{k}_z;x_2-x_2',y_2-y_2',z_2-z_2';t_1) 
\end{split}
\end{equation}

\item[$\boldsymbol{[z_2]}$] 
We apply the transmission function on the $x$ and $y$ axis given by  Eqs.~\eqref{txx} and~\eqref{tyy} respectively.  
Let us first consider the integrals along the $x$ axis. Using Eq.~\eqref{sxassumption1} we can simplify in Eq.~\eqref{rho_3d_bg1} (which is contained in Eq.~\eqref{rho_3d_bg}):
\begin{equation}\label{rho_3d_bg2}
\rho^{\text{\tiny QM(1)}}_2(x,x',x_1) t(x)t^*(x')= \exp \left[-\frac{
i( -x^2 + x'^2 -2  x_1 (x- x')) \frac{\sigma_2}{2 \sigma_1 }} 
{\sigma_2^2 }\right] t(x)t^*(x').
\end{equation}
The dependence on $\tilde{w}_x$, which will be integrated out, is contained in:
\begin{equation}\label{xfactors}
e^{-\frac{i}{\hbar}\tilde{k}_x \tilde{w}_x} 
\rho^{\text{\tiny QM(1)}}_2(x_2+\tilde{w}_x;x_2'+\tilde{w}_x) = \exp \left[
\frac{iB(x_2,x_2')\tilde{w}_x}{\sigma_2^2}\right] Exp\left[\frac{ C(x_2,x_2')}{\sigma_2^2}\right],
\end{equation}
where 
\begin{align}
B(x_2,x_2') &= \frac{\sigma_2}{\sigma_1}(x_2-x_2')  - \frac{1}{\hbar} \tilde{k}_x \sigma_2^2 \\
C(x_2,x_2') &= \frac{i\sigma_2}{2\sigma_1} \left( (x_2^2-x_2'^2) + 2x_1 (x_2-x_2') \right).
\end{align}
Hence the $\tilde{w}_x$ integral yields the Dirac delta function $\delta(\tilde{k}_x-(x_2-x_2')\frac{m}{t_1})$, which we use to perform $\tilde{k}_x$ integration. On the $y$ axis, by assumption \eqref{syassumption1}, we can replace $t_y(y)$ by 1. To summarize, after performing the $x$-axis integrations we obtain: 
\begin{equation}\label{rho_3d_ag}
\begin{split}
\rho_2(x_2, y_2, z_2;x'_2, y'_2, z'_2) = 
&e^{\frac{ik}{2L_1}(x_2^2-x_2'^2)}e^{-\frac{ik}{L_1}(x_2-x_2')x_1}\\
& \times\int d\tilde{k}_y \int d\tilde{w}_z \rho^{\text{\tiny QM(1)}}_2(y_2 +\tilde{w}_y ,y'_2 +\tilde{w}_y;0) e^{-\frac{i}{\hbar}\tilde{k}_y \tilde{w}_y} \\
&\times \int d\tilde{k}_z \int d\tilde{w}_z \rho^{\text{\tiny QM(1)}}_2(z_2+\tilde{w}_z,z'_2+\tilde{w}_z;0) e^{-\frac{i}{\hbar}\tilde{k}_z \tilde{w}_z} \\
&\times F(\frac{\hbar k}{L_1}(x_2-x_2'),\tilde{k}_y,\tilde{k}_z;x_2-x_2',y_2-y_2',z_2-z_2';t_1) 
\end{split}
\end{equation}

\item[$\boldsymbol{[z_2}$ to $\boldsymbol{z_3]}$] 
We apply Eq.~\eqref{rho_3d} to $\rho_2(x_2, y_2, z_2;x'_2, y'_2, z'_2)$ for a time $t_2=\frac{L_2}{v}$:
\begin{equation}\label{rho_3d_d}
\begin{split}
\rho_3(x_3, y_3, z_3;x'_3, y'_3, z'_3) = & \int d\tilde{\tilde{k}}_x \int d\tilde{\tilde{w}}_x  e^{-\frac{i}{\hbar}\tilde{\tilde{k}}_x \tilde{\tilde{w}}_x}  
\int d\tilde{\tilde{k}}_y \int d\tilde{\tilde{w}}_z  e^{-\frac{i}{\hbar}\tilde{\tilde{k}}_y \tilde{\tilde{w}}_y} \\
&\times\int d\tilde{\tilde{k}}_z \int d\tilde{\tilde{w}}_z  e^{-\frac{i}{\hbar}\tilde{\tilde{k}}_z \tilde{\tilde{w}}_z} 
F(\tilde{\tilde{k}}_x,\tilde{\tilde{k}}_y,\tilde{\tilde{k}}_z;x_3-x_3',y_3-y_3',z_3-z_3';t_2) \\
& \times \rho_3^{\text{\tiny QM}} (x_3 + \tilde{\tilde{w}}_x, y_3+ \tilde{\tilde{w}}_y, z_3+ \tilde{\tilde{w}}_z;x'_3+ \tilde{\tilde{w}}_x, y'_3+ \tilde{\tilde{w}}_y, z'_3+ \tilde{\tilde{w}}_z), 
\end{split}
\end{equation}
where
\begin{equation}\label{rho_3d_d_qm}
\begin{split}
\rho_3^{\text{\tiny QM}} (x_3, y_3, z_3;x'_3, y'_3, z'_3) =  & \int_{-\infty}^{+\infty} dx_2 \int_{-\infty}^{+\infty} dx_2' e^{\frac{ik}{2L_2}((x_3-x_2)^2-(x_3'-x'_2)^2)}  \\
&\times \int_{-\infty}^{+\infty} dy_2 \int_{-\infty}^{+\infty} dy_2' e^{\frac{ik}{2L_2}((y_3-y_2)^2-(y_3'-y'_2)^2)} \\
&\times \int_{-\infty}^{+\infty} dz_2 \int_{-\infty}^{+\infty} dz_2' e^{\frac{ik}{2L_2}((z_3-z_2)^2-(z_3'-z'_2)^2)} \\
&\times\rho_2(x_2, y_2, z_2;x'_2, y'_2, z'_2)
\end{split}
\end{equation}
is the free quantum mechanical evolution of $\rho_2(x_2, y_2, z_2;x'_2, y'_2, z'_2)$ for a time $t_2$.

\item[$\boldsymbol{[z_3]}$] 
We now set $x=x_3=x_3'$, $y=y_3=y_3'$ and $z=z_3=z_3'$ to obtain the probability density function
$p_3(x, y, z)= \rho_3(x, y, z;x, y, z)$, as the molecule interacts with the detector. However, we are only interested in the probability of detecting a particle at a horizontal coordinate $x$, therefore we consider: 
\begin{equation}\label{p3_3d}
p(x) = \int_{-\infty}^{+\infty} dy \int_{-\infty}^{+\infty} dz \;p_3(x, y, z).
\end{equation}

It is straightforward to perform the integrations along the $x$ axis in Eq.~\eqref{p3_3d} at this step of the calculation. In fact, these calculations are completely analogous to those described above (Eqs. \eqref{grwdp1} to \eqref{Dfunc_sep}), when working within the 1D approximation.

Let us now look at the tedious integrations associated with the $y$ axis in Eq.~\eqref{p3_3d}. In particular, we have from Eq.~\eqref{rho_3d_d}:
\begin{equation}
\int dy \int dy_2 \int dy_2'
\rho_2(x_2,y_2,z_2;x_2',y_2',z_2') 
e^{\frac{ik}{2L_2}(-2(y+\tilde{\tilde{w}}_y)y_2+2(y+\tilde{\tilde{w}}_y)y_2')} 
 e^{\frac{ik}{2L_2}(y_2^2-y_2'^2)}.
\label{paexp}
\end{equation}
By performing the $y$ integration we obtain a Dirac delta function $\delta(y_2-y_2')$ and  by performing then also the $y_2'$ integration, the expression given in Eq.~\eqref{paexp} reduces to:
\begin{equation} 
\int dy_2 \rho_2(x_2,y_2,z_2;x_2',y_2,z_2').
\label{calculation_z_2_ind_w}
\end{equation}
Let us now write the integrations associated with the $y$ axis contained within $\rho_2$ (see Eqs.~\eqref{rho_3d_ag} and~\eqref{rho_3d_bg1}):
\begin{equation}
\begin{split}
\int d\tilde{k}_y \int d\tilde{w}_y  \int dy_2 
&Exp \left[
-\frac{2\tilde{w}_y^2 +2 y_2^2 
+ \tilde{w}_y (4 y_2+\frac{i}{\hbar} \tilde{k}_y \sigma_2^2)} 
{\sigma_2^2 }
\right] \\
&\times F(\tilde{k}_x,\tilde{k}_y,\tilde{k}_z; x_2-x_2', 0,z_2-z_2';t_1).
\end{split}
\end{equation}
By performing the $y_2$ integration we remove the quadratic term containing $\tilde{w}_y$:
\begin{equation}\label{pa2exp}
\int d\tilde{k}_y \int d\tilde{w}_y F(\tilde{k}_x,
\tilde{k}_y,\tilde{k}_z; x_2-x_2', 0,z_2-z_2';t_1) 
 \exp \left[
-\frac{i}{\hbar} \tilde{w}_y \tilde{k}_y
\right].
\end{equation}
The $\tilde{w}_y$ integration yields a Dirac delta function $\delta(\tilde{k}_y)$ and by then also performing the $\tilde{k}_y$ integration, the expression given in Eq.~\eqref{pa2exp} reduces to:
\begin{equation}
 F(\tilde{k}_x,0,\tilde{k}_z; x_2-x_2', 0, z_2-z_2';t_1).
\end{equation}
In addition, we have just shown that $\rho_3^{QM}(x_3+\tilde{\tilde{w}}_x,y_3+\tilde{\tilde{w}}_y,z_3+\tilde{\tilde{w}}_z;x_3+\tilde{\tilde{w}}_x,
y_3+\tilde{\tilde{w}}_y,z_3+\tilde{\tilde{w}}_z)$ defined in Eq.~\eqref{rho_3d_d_qm} does not depend on $\tilde{\tilde{w}}_y$. Hence we can perform the following integrations: 
\begin{equation}\label{second_F_expr}
\int d\tilde{\tilde{w}}_y \int d\tilde{\tilde{k}}_y e^{-\frac{i}{\hbar}\tilde{\tilde{w}}_z \tilde{\tilde{k}}_z} F(\tilde{\tilde{k}}_x,\tilde{\tilde{k}}_y,\tilde{\tilde{k}}_z;0,0,0,t_2).
\end{equation}
Since the $\tilde{\tilde{w}}_y$ integration yields a Dirac delta function $\delta(\tilde{\tilde{k}}_y)$ we obtain from the expression given in Eq.~\eqref{second_F_expr}:
\begin{equation}
F(\tilde{\tilde{k}}_x,0,\tilde{\tilde{k}}_z;0,0,0,t_2).
\end{equation} 
We have thus shown that the final probability density is not affected by the  dynamics along the $y$ axis. A completely analogous calculation can be performed for the integrations associated with the $z$ axis. Hence we obtain from Eq.~\eqref{p3_3d}, relabeling $x_3$ as $x$, the interference pattern in Eq.~\eqref{p3_collapse}.
\end{description}

This calculation thus justifies, and gives the limits of applicability, of the 1D treatment discussed before.\\ \\

\begin{figure}[!htb]
\centering
\begin{minipage}[t]{0.5\textwidth}
  \includegraphics[width=1.0\linewidth]{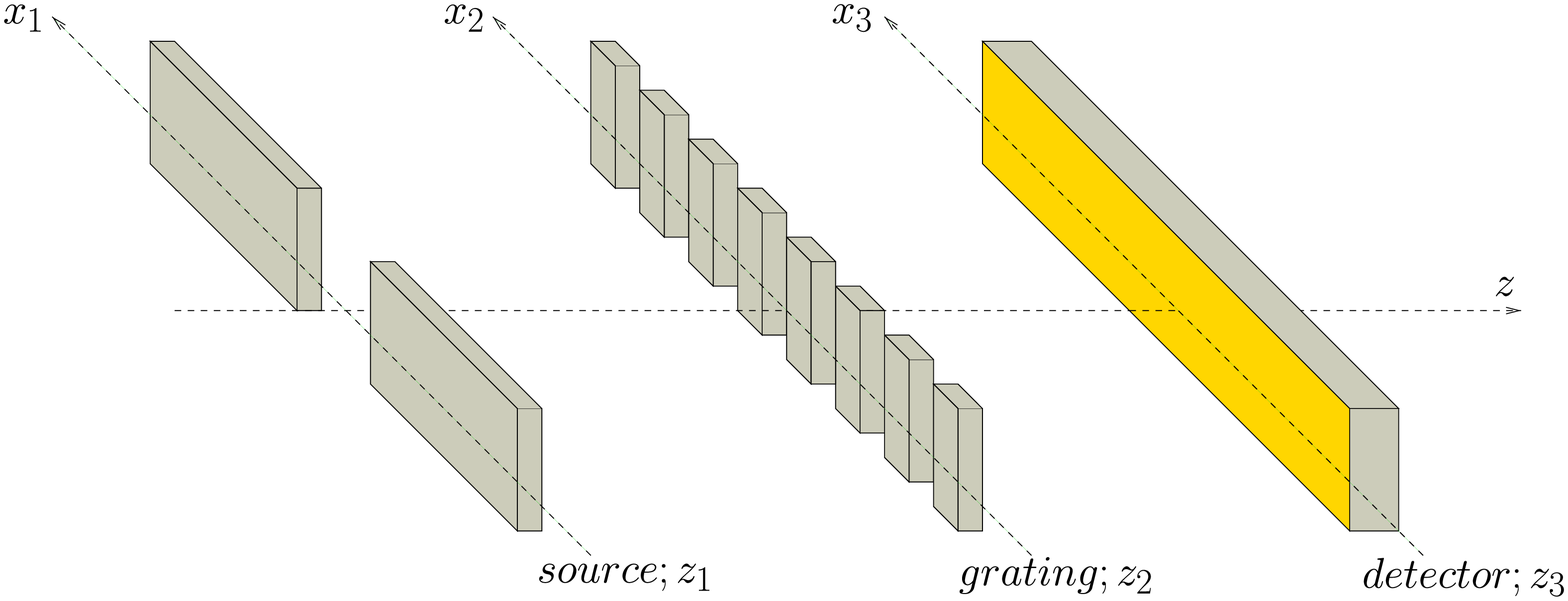}
\end{minipage}%
\begin{minipage}[t]{0.5\textwidth}
	\centering
  \includegraphics[width=1.0\linewidth]{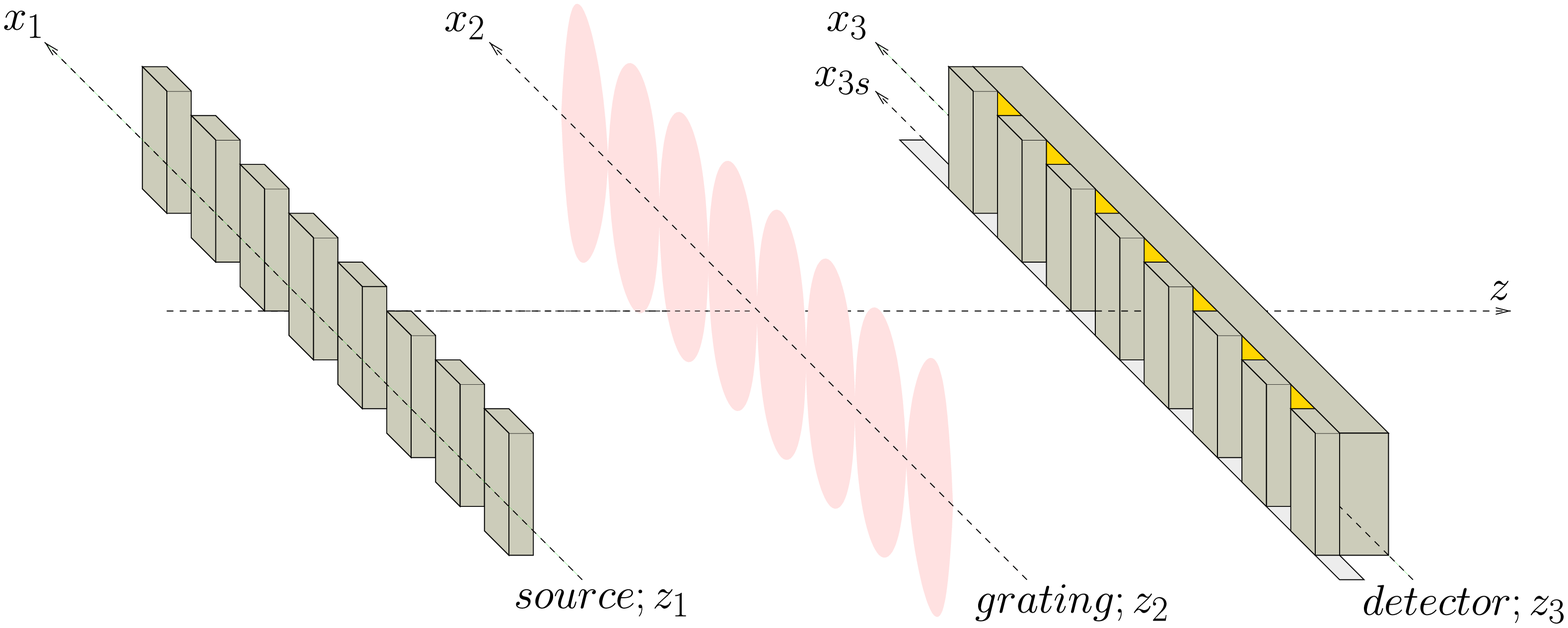}
\end{minipage}
\caption{\small Left: Far-field experimental setup. The optical elements are: an incoherent source at $z_1$ (centered on the optical axis, i.e. around $x_1=0$), the diffraction grating at $z_2$ (here we have shown a mechanical grating with $N=7$ slits) and the detector at $z_3$. For the experiment described in section \ref{desec} we have the following numerical values. The distance from  $z_1$ to $z_2$ is $L_1= 0.702\text{m}$ and the distance from $z_2$ to $z_3$ is $L_2=0.564\text{m}$. The source extension is taken to be $s=1 \mu\text{m}$. The  mechanical grating with $N=30$ slits is described by the period $d=100\text{nm}$ and slit width $l=79 \text{nm}$. The van der Walls forces due to the grating are modelled by an effective slit width $l_{eff}=43\text{nm}$. Right: Talbot Lau near-field experimental setup. In this case the optical elements are: an extended incoherent source at $z_1$, a diffraction grating at $z_2$ (here, an optical grating produced by a standing light wave) and the detector at $z_3$. Two additional mechanical gratings block part of the molecules: the mechanical grating located immediately after the source is held fixed, while the mechanical grating immediately before the detector can move along the $x_3$ axis (we denote the displacement from its initial position by $x_{3s}$). We assume that all elements have a very large horizontal extension such that one can approximate them with periodic functions. The detector at $z_3$ records molecules that arrive at all points along the $x_3$ axis in a certain amount of time. For the experiment described in section \ref{desec} we have the following numerical values. The distance from  $z_1$ to $z_2$ and the distance from $z_2$ to $z_3$ is $L=L_1=L_2=10.5\text{cm}$. Both mechanical gratings are described by the same period $d=266\text{nm}$ and slit width $l=110\text{nm}$. The optical grating is described by the wavelength $\lambda_{laser}=532\text{nm}$, the laser power $P_{laser}=1\text{W}$, the optical polarizability $\alpha_{opt}=410 \AA \times 4\pi\epsilon_0$ and the absorption cross section $\sigma_a=1.7 \times 10^{-21}\text{m}^2$.}
\label{fig:experimental_setups}
\end{figure}

\noindent {\bf Far-field}.
The experimental setup for the far-field interference experiments is summarized in Fig.~\ref{fig:experimental_setups} (left). 
The difference with respect to the idealized situation described here above is that instead of a single point source we have an incoherent source of horizontal extension $s$, centred at  $x_1=0$. We obtain the interference pattern by integrating Eq.~\eqref{p3_collapse} over the points $x_1$ of the source from $-\frac{s}{2}$ to $\frac{s}{2}$:
\begin{equation}\label{p3_far}
\begin{split}
p(x)=&\int_{-\infty}^{+\infty} dx_2\int_{-\infty}^{+\infty}  dx_2' 
  D(x_2-x_2') e^{-\frac{ik}{L_2}(x_2-x_2')x}\\ 
  &\times \sinc \left(\frac{k}{2L_1}(x_2-x_2')s \right) e^{ik(x_2^2-x_2'^2)(\frac{1}{2L_1}+\frac{1}{2L_2})},
\end{split}
\end{equation}
A related study of far-field decoherence effects in the Wigner function formalism is given in \cite{PhysRevA.73.052102}.

Let us discuss how to evaluate numerically Eq.~\eqref{p3_far}. We recognize from the factor $e^{-\frac{ik}{L_2}(x_2-x_2')x_3}$  a Fourier transform and an inverse Fourier transform. Fourier transforms can be approximated with discrete Fourier transforms using the FFT algorithm. Hence the integrations in Eq.~\eqref{p3_far} can be conveniently  evaluated numerically with the row column FFT algorithm. \\ \\

\noindent {\bf Talbot Lau near-field}.
The experimental setup for the KDTL near-field interference experiment is represented in Fig.~\ref{fig:experimental_setups} (right). This is essentially the same scheme as presented before (Fig.~\ref{fig_paraaxial}) except that now we have two additional gratings at positions $z_1$, $z_3$ along the optical axis. We assume that all gratings have a very large horizontal extension such that we can model them by  periodic functions. The first grating at $z_1$ acts as a mask of an infinite incoherent source and similarly the third grating at $z_3$ acts as a mask of the infinite detection screen. The experiment is performed by moving the masking grating at $z_3$ along the $x_3$ axis and recording the total number of molecules that reach the detector in a given amount of time. At the end one obtains the number of molecules that reach the detector given a given displacement $x_{3s}$ of the third grating from its initial position.

In section \ref{desec} we will describe the KDTL experiment, where the 3 gratings have the same periodicity $d$ and the distance from $z_1$ to $z_2$ and from $z_2$ to $z_3$ is $L=L_1=L_2$. Due to the periodicity of the 3 gratings, we adopt the following notation for the Fourier series of the corresponding transmission functions (notation of Ref.~\cite{1464-4266-5-2-362}):
\begin{align}
|t_1(x_1)|^2 &= \sum_{l=-\infty}^{+\infty} A_l e^{ i2\pi l\frac{x_1}{d}}, \label{t1}\\
t(x_2) &= \sum_{j=-\infty}^{+\infty} b_j e^{ i2\pi j\frac{x_2}{d}}, \label{t2}\\
|t_3(x_3)|^2 &= \sum_{n=-\infty}^{+\infty} C_n e^{i2\pi n\frac{x_3}{d} } . \label{t3}
\end{align}
We can now directly proceed with the derivation of the interference pattern starting again from Eq.\eqref{p3_collapse}:
\begin{equation}\label{sx3si}
S(x_{3s})=\int_{-\infty}^{+\infty} \int_{-\infty}^{+\infty} dx_1 dx_3 p(x_3; x_1)
|t_1(x_1)|^2 |t_3(x_3-x_{3s})|^2 ,
\end{equation}
where with respect to the far-field experiment, there is a further integration over all the detector region,  $x_{3s}$ is the horizontal shift of the third grating and $p(x_3; x_1)$ is the interference pattern due to a single source point at $x_1$ given by Eq.~\eqref{p3_collapse}. In other words, $S(x_{3s})$ gives the number of molecules that reach the infinite detector at $z_3$ from the infinite source at $z_1$ in a given amount of time, given a displacement $x_{3s}$ of the third grating from its initial position. Since this gives formally an infinite value, we have to properly normalize the result. This is done in the following way.

The integrations in Eq.~\eqref{sx3si} over $x_1$, $x_3$ yield two delta functions $\delta(\frac{2\pi l}{d} - \frac{k}{L_1}(x_2-x'_2))$, $\delta(\frac{2\pi n}{d} - \frac{k}{L_2}(x_2-x'_2))$. We perform the integration over $dx'_2$ which gives the constraint $x_2-x'_2=\frac{2\pi l}{d}\frac{L}{k}$, while the other delta function gives the constrain $l=n$. We now divide by $\delta(0)$ in order to remove the infinite factor due to the delta function giving this first constrain. We are left with the integration over $dx_2$ which gives a delta function $\delta(\frac{4\pi n}{d} + \frac{2\pi j}{d} - \frac{2\pi j'}{d})$, where $j'$ is the index in the Fourier expansion of $t^*(x'_2)$. This gives the constraint $j'=j+2n$.  We again divide by $\delta(0)$ in order to remove the infinite factor due to the delta function giving this second constrain. We are now left with a finite expression. In order to obtain a notation consistent with that of \cite{1464-4266-5-2-362} we relabel $n$ as $-n$ and use the fact that $A_{-n}=A^*_n$, $C_{-n}=C^*_n$. Thus we obtain:
\begin{equation}\label{p3_near}
S(x_{3s})=\sum_n A^*_n C^*_n B_n D\left(\frac{2\pi n }{d}\frac{L}{k}\right) e^{i 2 \pi n \frac{x_{3s}}{d}},
\end{equation}
where $B_n=\sum_j b_j b^*_{j-n} e^{i \frac{\pi^2}{d^2} \frac{L}{k}(n^2-2nj)}$.
The above equation coincides with the results derived by using the Wigner function formalism \cite{PhysRevA.83.043621,PhysRevA.70.053608}. 

\clearpage
\section{Summary of Collapse Models and of the interference pattern}\label{socm}

\noindent {\bf CSL: Continuous Spontaneous Localization}. Here we are referring to the mass-proportional version of the CSL model ~\cite{PhysRevLett.73.1}. The single-particle master equation in 3D is given by~\cite{PhysRevA.90.062105}:
\begin{equation}\label{csl_equation}
\frac{d }{d t}\hat{\rho}(t)= 
-\frac{i}{\hbar}\left[\hat{H},\hat{\rho}(t) \right] 
+\lambda \frac{m^2}{m_0^2}  \left( \left(\frac{r_C}{\sqrt{\pi}\hbar}\right)^3 
\int d^3 \bm{Q} e^{-\frac{\bm{Q}^2r_C^2}{\hbar^2}}
e^{\frac{i}{\hbar} \bm{Q} \cdot \hat{\bm{x}}}\hat{\rho}(t)e^{-\frac{i}{\hbar} \bm{Q} \cdot \hat{\bm{x}}}- \hat{\rho}(t)\right).
\end{equation}
The physical meaning of the phenomenological constants $\lambda$ and $r_C$ was clarified in section \ref{introduction}. In the free-particle case $\hat{H} = \hat{p}^2/2m$, the equation can be solved exactly. In the coordinate basis, it reads~\cite{PhysRevD.34.470}:
\begin{equation}\label{rho_csl}
\rho^{\text{\tiny CSL}}(\bm{x},\bm{x}',t)= \frac{1}{(2\pi\hbar)^3} \int d \bm{\tilde{k}}  \int \bm{\tilde{w}} e^{-\frac{i}{\hbar}\bm{\tilde{k}} \cdot \bm{\tilde{w}}} 
 F_{\text{\tiny CSL}}(\bm{\tilde{k}},\bm{x}-\bm{x}',t) \rho^{\text{\tiny QM}}(\bm{x}+\bm{\tilde{w}},\bm{x}'+\bm{\tilde{w}},t),
\end{equation}
where $\rho^{\text{\tiny QM}}(\bm{x},\bm{x}',t)$ is the standard free quantum evolution for the density matrix ($\lambda = 0$) and 
\begin{equation}\label{FF}
F_{\text{\tiny CSL}}(\bm{\tilde{k}},\bm{q},t)= \exp\left[-\lambda \frac{m^2}{m_0^2} t \left(1-\frac{1}{t}\int_0^t d\tau e^{-\frac{1}{4r_C^2}(\bm{q}-\frac{\bm{\tilde{k}}\tau}{m})^2} \right) \right].
\end{equation}
The interference pattern is given by Eq.~\eqref{p3_collapse} with the function $D$ defined as follows:
\begin{equation}\label{cfunc}
D_{\text{\tiny CSL}}(x_2-x_2')=\exp \left[-\lambda \frac{m^2}{m_0^2} (t_1+t_2)
\left(1-\frac{\sqrt{\pi}}{2}\frac{\erf (\frac{(x_2-x_2')}{2r_C})}{\frac{(x_2-x_2')}{2r_C}} \right) \right].
\end{equation}
Note that $D_{\text{\tiny CSL}}(x_2-x_2')$ was previuosly derived~\cite{PhysRevA.83.043621,PhysRevA.70.053608, PhysRevA.73.052102} by using the Wigner function's formalism. 

The GRW single-particle master equation has the same mathematical structure as the CSL single-particle master equation. Since our analysis is based entirely on this master equation the above CSL formulae apply also to the GRW model, the only difference being the amplification mechanism discussed before.  \\ \\

\noindent {\bf DP: Di\'{o}si-Penrose}.
\noindent The single-particle master equation in 3D for a particle
of mass $m_{0}$ is given by~\cite{PhysRevA.40.1165,PhysRevA.90.062135}:
\begin{equation}
\frac{d\hat{\rho}_{t}}{dt}=-\frac{i}{\hbar}\left[\hat{H},\hat{\rho}_{t}\right]+\frac{Gm_{0}^{2}}{2\pi\hbar^{2}}\int d\bm{Q}\frac{1}{\bm{Q}^{2}}e^{-\frac{\bm{Q}^{2}R_{0}^{2}}{\hbar^{2}}}\left(e^{\frac{i}{\hbar}\bm{Q}\cdot\hat{x}}\hat{\rho_{t}}e^{-\frac{i}{\hbar}\bm{Q}\cdot\hat{x}}-\hat{\rho_{t}}\right),\label{dp_equation}
\end{equation}
where $R_{0}$ is a regularization parameter, which has to be included
in order to avoid divergences at short distances~\footnote{The DP model introduces only one cut-off length phenomenological parameter
$R_{0}$, which cures the ultraviolet divergence of the gravitational
interaction. The effective collapse rate, analogous to $\lambda$,
is given by $Gm_{0}^{2}/\sqrt{\pi}\hbar R_{0}$, while $R_{0}$ describes
how well an object is localized, analogous to $r_{C}$.}. For a point-like particle~\footnote{We thank Prof. Lajos Di\'{o}si for signaling this point. }
of mass $m$ we have to replace $m_{0}$ with $m$. In the free-particle
case, the equation can be solved exactly, and in the position representation
it reads~\cite{PhysRevA.90.062135}: 
\begin{equation}\label{rho_grw_dp3}
\rho^{\text{\tiny DP}}(\bm{x},\bm{x'},t)= \frac{1}{(2\pi\hbar)^3} \int d\bm{\tilde{k}}  \int \bm{\tilde{w}} e^{-\frac{i}{\hbar}\bm{\tilde{k}} \cdot \bm{\tilde{w}}}
F_{\text{\tiny DP}}(\bm{\tilde{k}},\bm{x}-\bm{x'},t) \rho^{\text{\tiny QM}}(\bm{x}+\bm{w},\bm{x'}+\bm{w},t),
\end{equation}
where, again, $\rho^{\text{\tiny QM}}(\bm{x},\bm{x}',t)$ is the free standard quantum evolution, and
\begin{equation}\label{FF_dp3}
F_{\text{\tiny DP}}(\bm{\tilde{k}},\bm{q},t)= \exp\left[ -\frac{1}{\hbar}\\
\int_0^t d\tau \left(   U(-\frac{\bm{\tilde{k}} \tau}{m} +\bm{q}) - U(0) \right)  \right]
\end{equation}
with $U(\bm{x})=-G m_0^2 \erf(|\bm{x}|/2R_0)/|\bm{x}|$.

The interference pattern is given again by Eq.~\eqref{p3_collapse}, with the function $D$ given by:
\begin{equation}\label{cfunc_dp}
D_{\text{\tiny DP}}(x_2-x_2')= \exp \left[-\frac{G m_0^2 }{\hbar \sqrt{\pi} R_0} (t_1+t_2)  \left(1- {}_2F_2 \left(\frac{1}{2},\frac{1}{2},\frac{3}{2},\frac{3}{2};  -\left(\frac{|x_2-x_2'|}{2 R_0} \right)^2 \right) \right) \right],
\end{equation}
where ${}_2F_2(\frac{1}{2},\frac{1}{2},\frac{3}{2},\frac{3}{2}; z) = \sum_{k=0}^{\infty} \left( \frac{1}{1+2k}\right)^2 \frac{z^k}{k!}$.

\begin{figure}[!htb]
\begin{center}
\includegraphics[width=0.75\textwidth]{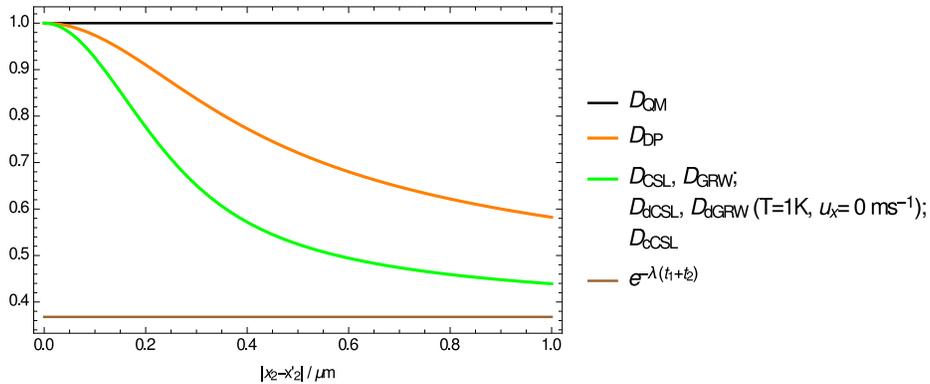}
\caption{Comparison of $D$ functions for the considered collapse models. The plot is obtained with $r_C=10^{-7}m$, $\lambda=500 s^{-1}$, $t_1=t_2=1ms$, $L_1=L_2=0.1m$, $R_0=10^{-7}m$, $\lambda_{\text{\tiny DP}}=\lambda=500 s^{-1}$, where the rescaled $\lambda$, $\lambda_{DP}$ are such that $\lambda (t_1+t_2)=1$.  The black solid line represents the quantum mechanical function ($D=1$), the orange solid line represents the $D$ function for the DP model, the green solid line represents that of the  CSL, GRW, dCSL, dGRW and cCSL models (for temperatures $T > 10^{-7}\text{K}$ and boost along the $x$ axis $u_x < 10^{4}\text{ms}^{-1}$ . The solid brown line represents the asymptotic value of the $D$ functions for all the considered collapse models as $|x-x'| \rightarrow +\infty$.}
\label{DPvsCSL}
\end{center}
\end{figure}

It is instructive to compare $D_{\text{\tiny DP}}$ and $D_{\text{\tiny CSL}}$. One can relate the role of $\lambda$ in the CSL model with $\lambda_{\text{\tiny DP}}=\frac{G m_0^2}{\hbar \sqrt{\pi} R_0}$ in the DP model, and the role of $r_C$ for CSL with $R_0$ for DP. As Fig.~\ref{DPvsCSL} shows, when appropriately rescaled, $D_{\text{\tiny DP}}$ and $D_{\text{\tiny CSL}}$ have a very similar behaviour. In particular, both are equal to $1$ for  $|x_2-x_2'|=0$ and decay more or less in the same way towards the asymptotic value $e^{-\lambda (t_1+t_2)}$ as $|x_2-x_2'| \rightarrow \infty$. \\ \\

\noindent {\bf dCSL: Dissipative CSL.} This is a recently developed new version of the CSL model, which includes dissipative effects, which prevent the energy of the system to increase and eventually diverge.  The single-particle master equation  in 3D is~\cite{Smirne:2014paa,PhysRevA.90.062135}:
\begin{equation}\label{csl_equation_dcsl}
\begin{split}
&\frac{d \hat{\rho}_t}{dt} =  -\frac{i}{\hbar}\left[ \hat{H}, \hat{\rho}_t\right] \\
&+\lambda \frac{m^2}{m_0^2} \left( \left(\frac{r_C(1+k_T) }{\sqrt{\pi} \hbar}\right)^3 
\int d \bm{Q} e^{\frac{i}{\hbar} \bm{Q} \cdot \bm{\hat{x}} }  e^{-\frac{r_C^2}{2\hbar^2}((1+k_T)\bm{Q}+2k_T\bm{\hat{p}})^2 } \hat{\rho_t}
 e^{-\frac{r_C^2}{2\hbar^2}((1+k_T)\bm{Q}+2k_T\bm{\hat{p}})^2 } e^{-\frac{i}{\hbar} \bm{Q}\cdot \bm{\hat{x}} } - \hat{\rho} \right)
\end{split}
\end{equation}
where $k_T=\frac{\hbar^2}{8 m r_C^2 k_B T}$, $k_B$ is the Boltzmann constant and $T$ the temperature the system thermalizes to. This is a new parameter of the theory, which together with $\lambda$ and $r_C$ fully identifies the model. In the limit $k_T \rightarrow 0$ (i.e $T \rightarrow \infty$), one re-obtains standard CSL model. 

\begin{figure}[!htb]
\centering{}\includegraphics[width=0.5\textwidth]{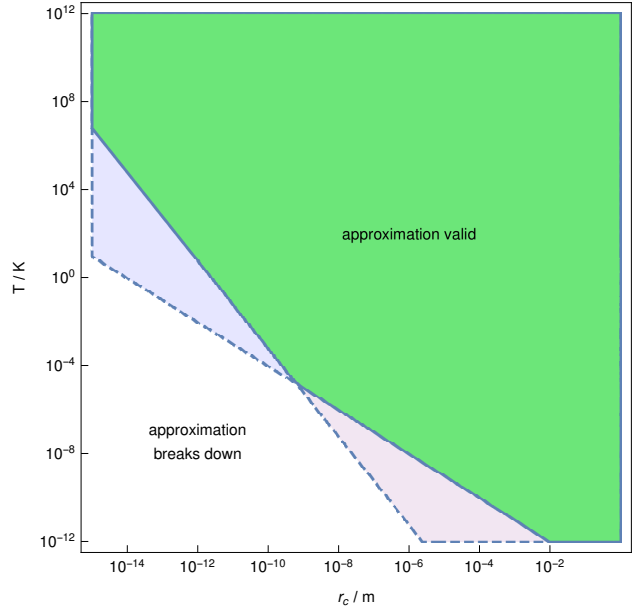}
\caption[Limits of validity (temperature)]{ Graphical depiction of the conditions given in Eqs. (\ref{ktcond}),
(\ref{eq:dcsl_cond1}). The condition given by Eq. (\ref{ktcond})
is satisfied in the orange and green regions, while the condition
given in Eq. (\ref{eq:dcsl_cond1}) is satisfied in the gray and green
regions: both conditions are satisfied in the green region. We estimate these limits for experimental situations, where the spatial and temporal extension of the superpositions is limited to distances $\Delta x<10^{-5}\text{m}$ and duration $t<10^{-2}\text{s}$, respectively.}
\label{limitsT} 
\end{figure}

We simplify the analysis, as in \cite{PhysRevA.90.062135,Smirne:2014paa}, by considering only small values of $k_T$:
\begin{equation}\label{ktcond}
k_T \ll 1.
\end{equation}
This assumption identifies a region in the parameter space $(T,r_C)$, depicted in figure \ref{limitsT}.

In the free particle case the solution reads~\cite{Smirne:2014paa}:
\begin{equation}\label{rho_grw_dcsl}
\rho^{\text{\tiny dCSL}}(\bm{x},\bm{x}',t)= \frac{1}{(2\pi\hbar)^3} \int d\bm{\tilde{k}}  \int \bm{\tilde{w}} e^{-\frac{i}{\hbar}\bm{\tilde{k}} \cdot \bm{\tilde{w}}} F_{\text{ \tiny dCSL}}(\bm{\tilde{k}},\bm{x}-\bm{x}',t) \rho^{\text{\tiny QM}}(\bm{x}+\bm{\tilde{w}},\bm{x}'+\bm{\tilde{w}},t),
\end{equation}
where as usual $\rho^{\text{\tiny QM}}(\bm{x},\bm{x}',t)$ is the free standard quantum evolution, and
\begin{equation}\label{FF_dcsl}
F_{\text{\tiny dCSL}}(\bm{\tilde{k}},\bm{q},t)=  \exp\left[ -\lambda \frac{m^2}{m_0^2} t  \left(1-\frac{1}{t}\int_0^t d\tau e^{-\frac{\bm{\tilde{k}}^2r_C^2 k_T^2 }{\hbar^2} -\frac{(-\frac{\bm{\tilde{k}}\tau}{m}+\bm{q})^2}{4r_C^2 (1+k_T)^2} } \right) \right].
\end{equation}

The interference pattern is still given by Eq.~\eqref{p3_collapse}, with the function $D$ given by:
\begin{equation}\label{cfunc_dcsl}
\begin{split}
D_{\text{\tiny dCSL}}(x_2-x_2')= &\exp \left[\right. -\lambda \frac{m^2}{m_0^2} (t_1+t_2) \\
&+ \lambda \frac{m^2}{m_0^2} \left( t_1 e^{-\frac{k^2}{L_1^2}(x_2-x'_2)^2 r_C^2 k_T^2}+t_2 e^{-\frac{k^2}{L_2^2}(x_2-x'_2)^2 r_C^2 k_T^2 }\right) 
\frac{\sqrt{\pi}}{2}\frac{\erf (\frac{(x_2-x_2')}{2 r_C(1+k_T)})}{\frac{(x_2-x_2')}{2r_C(1 +k_T)}} \left.\right].
\end{split}
\end{equation}

We note that this equation reduces to the CSL D-function, given in
Eq.~(\ref{cfunc}), when the following condition is fulfilled:
\begin{equation}
r_{C}t \gg \frac{\hbar\Delta x}{8k_{B}T},\label{eq:dcsl_cond1}
\end{equation}
for $t=t_{1}$ and $t=t_{2}$. 
We estimate these limits for experimental situations, where the spatial and temporal extension of the superpositions is limited to distances $\Delta x<10^{-5}\text{m}$
and duration $t<10^{-2}\text{s}$, respectively. This condition identifies a region in the parameter space $(r_{C},T)$, depicted in Fig. \ref{limitsT}.

\begin{figure}[!htb]
\centering{}\includegraphics[width=0.65\textwidth]{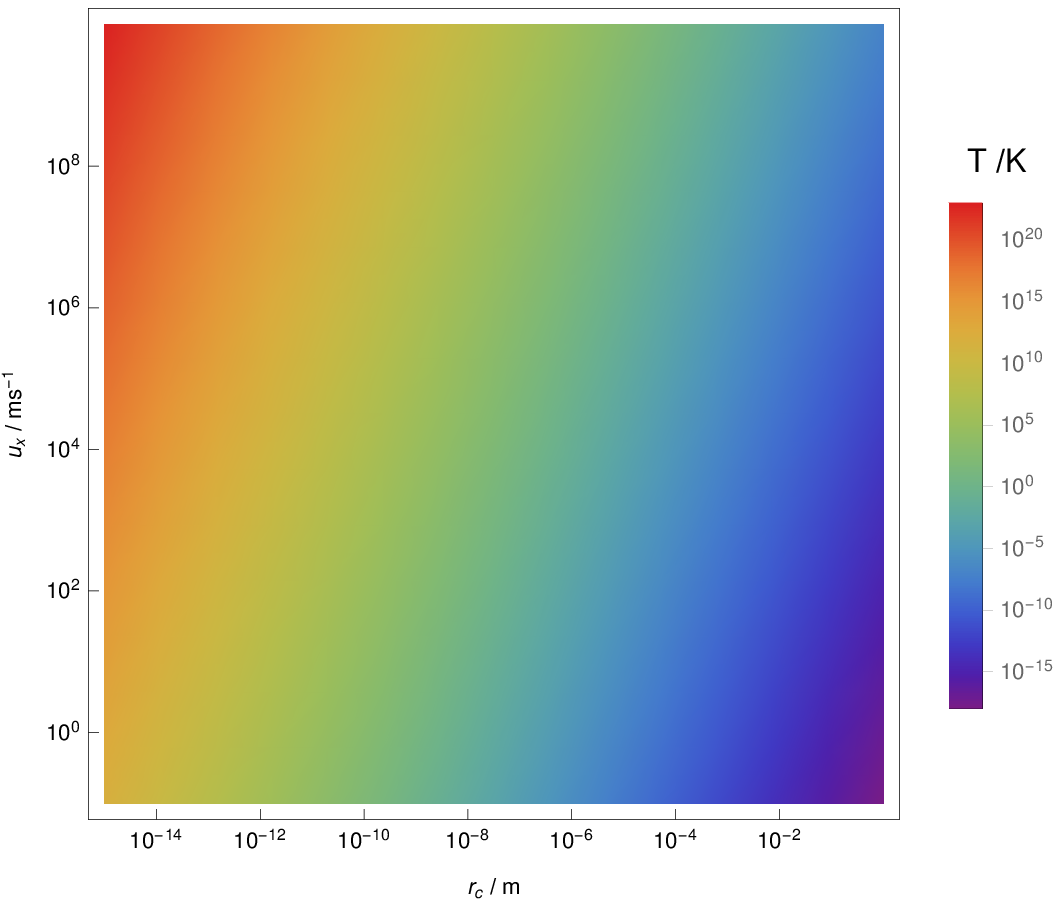}
\caption[Limits of validity (temperature and boosts)]{ Graphical depiction of the condition given in Eq. (\ref{eq:dcsl_cond2}).
The color indicates the minimum temperature, for a given value of
$r_{C}$ and $u_x$, such that the condition given in Eq. (\ref{eq:dcsl_cond2})
is satisfied. We estimate these limits for experimental situations, where the spatial and temporal extension of the superpositions is limited to distances $\Delta x<10^{-5}\text{m}$
and duration $t<10^{-2}\text{s}$, respectively.}
\label{limitsTB} 
\end{figure}

However, the master equation \eqref{csl_equation_dcsl} is not invariant under boosts. Indeed, the dissipative CSL master equation has the same structure of a  quantum linear Boltzmann equation of a particle immersed in a finite temperature bath \cite{Vacchini200971}. Thus the dissipative CSL model contains an additional free parameter, a velocity $\bm{u}$, which is analogous to the relative velocity between bath and particle. In particular, the master equation in the boosted reference frame with boost velocity $\bm{u}$ is given by the following equation:
\begin{equation}\label{csl_equation_dcsl_boost}
\begin{split}
\frac{d \hat{\rho}_t}{dt} = &-\frac{i}{\hbar}\left[ \hat{H}, \hat{\rho}_t\right]
+\lambda \frac{m^2}{m_0^2}\left(\right. \left(\frac{r_C(1+k_T) }{\sqrt{\pi} \hbar} \right)^3 \\
&\times \int d\bm{Q} e^{\frac{i}{\hbar} \bm{Q} \cdot \bm{\hat{X}} }  e^{-\frac{r_C^2}{2\hbar^2}((1+k_T) \bm{Q}+2k_T(\bm{\hat{P}} - m  \bm{u}))^2 } \hat{\rho_t} e^{-\frac{r_C^2}{2\hbar^2}((1+k_T)\bm{Q}+2k_T(\bm{\hat{P}} - m \bm{u}))^2 } e^{-\frac{i}{\hbar} \bm{Q} \cdot \bm{\hat{X}} } - \hat{\rho} \left.\right).
\end{split}
\end{equation}
We find the solution of Eq.~\eqref{csl_equation_dcsl_boost} using the characteristic function approach \cite{PhysRevA.82.042111}. The solution is given by Eq.~\eqref{rho_grw_dcsl} with the function $F_{\text{\tiny dCSL}}$ replaced by:
\begin{equation}\label{FF_dcsl_boost}
F_{\text{\tiny dCSL}}^{\text{\tiny boosted}}(\bm{\tilde{k}},\bm{q},t; \bm{u})= \exp\Bigg[-\lambda \frac{m^2}{m_0^2} t  \left(1-\frac{1}{t}\int_0^t d\tau e^{-\frac{\bm{\tilde{k}}^2r_C^2 k_T^2 }{\hbar^2} 
-\frac{(-\frac{\bm{\tilde{k}}\tau}{m}+\bm{q})^2}{4r_C^2 (1+k_T)^2}}
e^{\frac{i}{\hbar}\frac{2k_T m \bm{u}}{1+k_T}\cdot(-\frac{\bm{\tilde{k}}\tau}{m}+\bm{q}) }  \right) \Bigg].
\end{equation}
The interference pattern is given by Eq.~\eqref{p3_collapse} with the function $D$ replaced by:
\begin{equation}\label{cfunc_dcsl_boost}
\begin{split}
D_{\text{\tiny dCSL}}^{\text{\tiny boosted}}(x_2-x_2')= \exp \left[\right.&-\lambda \frac{m^2}{m_0^2} (t_1+t_2) \\
&+ \lambda \frac{m^2}{m_0^2} \left( t_1 e^{-\frac{k^2}{L_1^2}(x_2-x'_2)^2 r_C^2 k_T^2}+t_2 e^{-\frac{k^2}{L_2^2}(x_2-x'_2)^2 r_C^2 k_T^2 }\right) \\
&\times\frac{\int_0^{ (\frac{(x_2-x_2')}{2r_C(1+k_T)})}d\tau e^{-\tau^2}\cos(2\tau\frac{2 r_Ck_T m u_x}{\hbar} )}{ (\frac{ (x_2-x_2')}{2 r_C(1+k_T)})} \left. \right],
\end{split}
\end{equation}
where $u_{x}$ is the $x$ component of $\bm{u}$. We note that this
equation reduces to the CSL D-function, given in Eq. (\ref{cfunc}),
when in addition to Eq. (\ref{eq:dcsl_cond1}), the following condition
is fulfilled:
\begin{equation}
\frac{r_{C}^{2}}{u_{x}}\gg\frac{\hbar\Delta x}{8k_{B}T}\label{eq:dcsl_cond2}
\end{equation}
We note that Eq.~\eqref{cfunc_dcsl_boost} reduces to Eq.~\eqref{cfunc_dcsl}
as $u_{x}\rightarrow0$ and that Eq.~\eqref{cfunc_dcsl}
reduces to Eq.~\eqref{cfunc} as $k_{T}\rightarrow0$. We estimate these limits for experimental situations, where the spatial and temporal extension of the superpositions is limited to distances $\Delta x<10^{-5}\text{m}$ and duration $t<10^{-2}\text{s}$, respectively. This condition
identifies a region in the parameter space $(r_{C},u_{x},T)$, depicted
in Fig. \ref{limitsTB}.

\begin{figure}[!htb]
\includegraphics[width=0.75\textwidth]{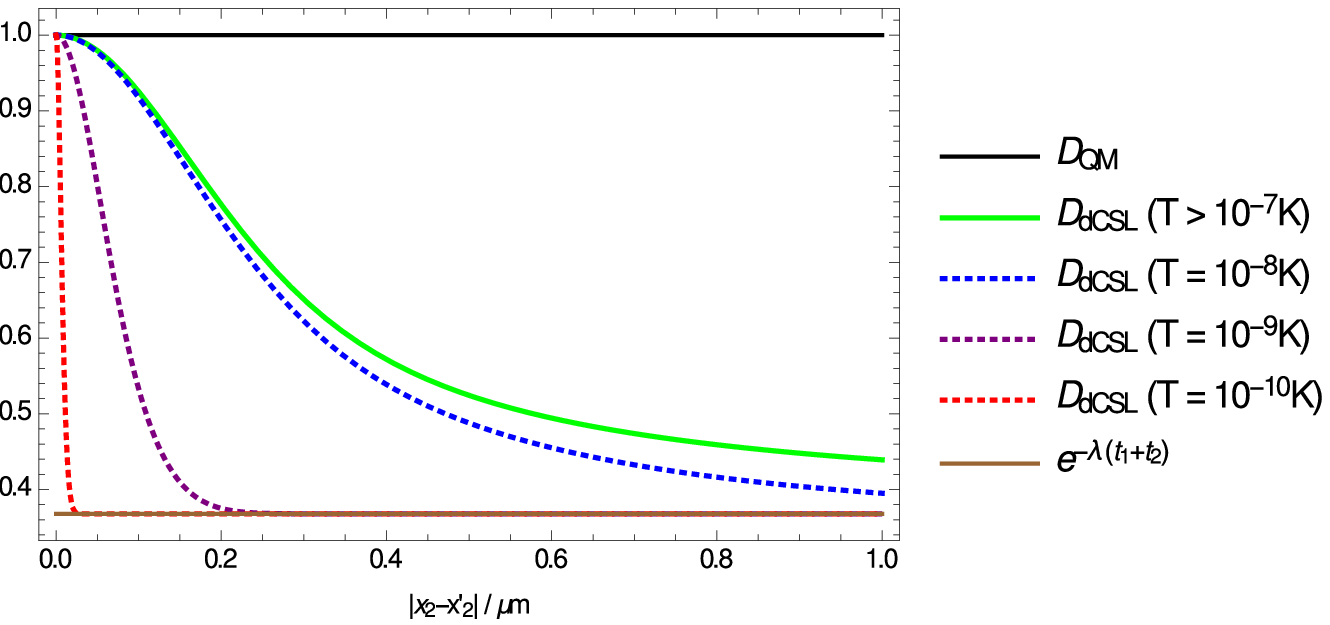}
\caption{Comparison of $D_{\text{\tiny dCSL}}$ functions for different temperatures $T$ at fixed boost $u_x=0$. The plot is obtained with $r_C=10^{-7}\text{m}$, $\lambda=500 \text{s}^{-1}$, $t_1=t_2=1\text{ms}$ and $L_1=L_2=0.1\text{m}$. The black solid line represents the quantum mechanical function ($D=1$), the green solid line represents the $D$ function for the dCSL models with $T>10^{-7}\text{K}$ (which includes the CSL model), while the dashed lines represent the dCSL models with temperatures $T=10^{-8}\text{K}$, $T=10^{-9}\text{K}$ and $T=10^{-10}\text{K}$. The solid brown line represents the asymptotic value of the $D$ functions for all the considered collapse models as $|x-x'| \rightarrow +\infty$.}\label{d_dCSL_temp}
\end{figure}

\begin{figure}[!htb]
\includegraphics[width=0.75\textwidth]{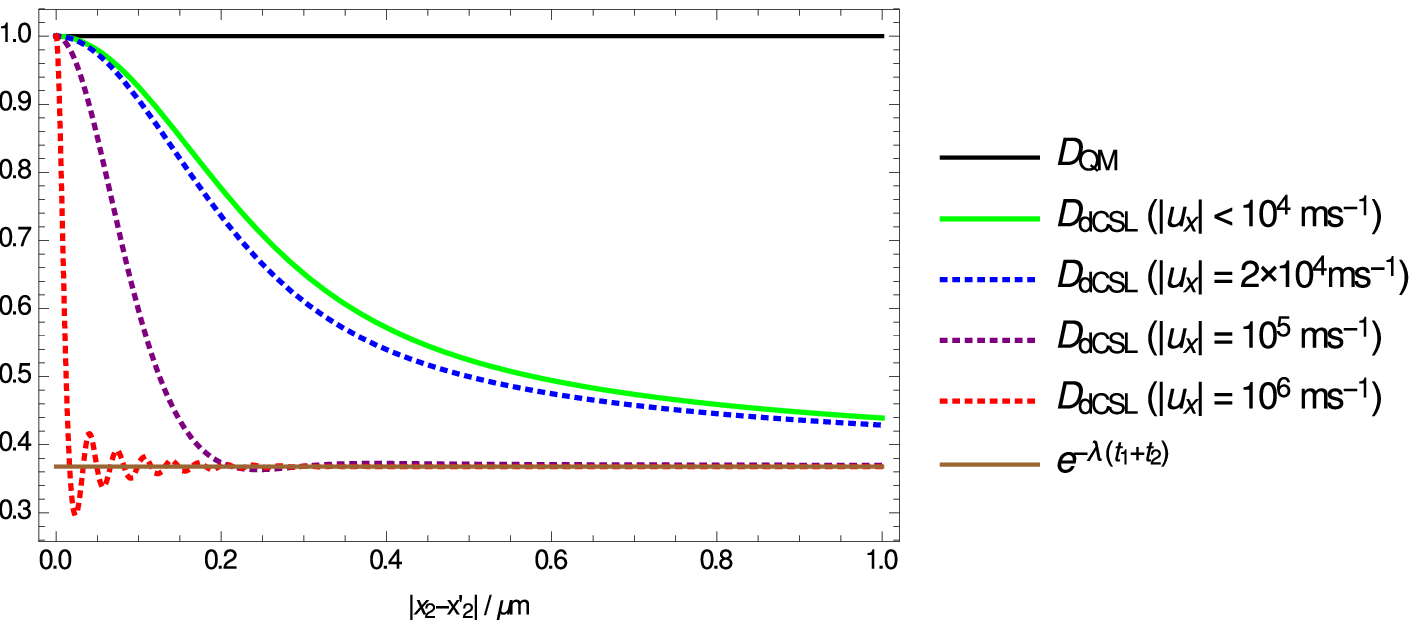}
\caption{Comparison of $D_{\text{\tiny dCSL}}$ functions for different boost along the $x$ axis $u_x$ at fixed temperature $T=1K$. The plot is obtained with $r_C=10^{-7}\text{m}$, $\lambda=500 \text{s}^{-1}$, $t_1=t_2=1\text{ms}$ and $L_1=L_2=0.1\text{m}$. The black solid line represents the quantum mechanical function ($D=1$), the green solid line represents the $D$ function for the dCSL models with boosts along the $x$ axis $|u_x| < 10^{4}\text{ms}^{-1}$ (which includes the CSL model), while the dashed lines represent the dCSL models with boost along the $x$ axis $|u_x| = 2\times 10^{4}\text{ms}^{-1}$, $|u_x| = 10^{5}\text{ms}^{-1}$ and $|u_x| = 10^{6}\text{ms}^{-1}$. The solid brown line represents the asymptotic value of the $D$ functions for all the considered collapse models as $|x-x'| \rightarrow +\infty$.}
\label{d_dCSL_boost}
\end{figure}

A comparison of $D_{\text{\tiny dCSL}}^{\text{\tiny boosted}}$ functions evaluated with different temperatures $T$ and different boosts $u_x$ is given in Figs.~\ref{d_dCSL_temp},~\ref{d_dCSL_boost} respectively. We see from these figures that the dCSL model with large temperatures $T$ and small boosts $u_x$ give the smallest modification with respect to the standard quantum mechanical evolution ($D=1$) and practically coincide with the CSL model evolution. Hence, given that $T$ and $u_x$ are unknown, the CSL model can be used as a bound for all dCSL models with arbitrary $T$ and $\bm{u}$.

The dGRW single-particle master equation has the same mathematical structure as the dCSL single-particle master equation. Since our analysis is based entirely on this master equation the above dCSL formulae apply also to the dGRW model, the only difference being the amplification mechanism discussed in section \ref{sec_amp}.\\ \\

\noindent {\bf cCSL: Colored CSL.}
This model presents an additional difficulty with respect to the white noise models discussed here above. The previous calculation splits into two parts, the free evolution from  the source at time $\tau_1$ to the grating at time $\tau_2$ and the free evolution from the grating at time $\tau_2$ to the detector at time $\tau_3$, whereas at time $\tau_2$ the molecule is subject to a non free evolution. Let us consider the times $\tau_1<t_{\text{\tiny before}}<\tau_2$ and $\tau_2<t_{\text{\tiny after}}<\tau_3$. The non white noise might correlate the evolution between $t_{\text{\tiny before}}$ and $t_{\text{\tiny after}}$. In order to simplify the analysis we neglect the correlations between these times by assuming a small correlation time $\tau_C \ll \tau_3-\tau_1$. A similar argument can be put forward for the correlation between times before and after $\tau_1$. Hence we limit the discussion to non white CSL models with small correlation times. In particular, this assumption justifies the following approximation of the free one particle master equation in 3D \cite{1751-8121-40-50-012,Donadi201470}:
\begin{equation}\label{nw_equation_momentum}
\begin{split}
\frac{d \hat{\rho}_t}{dt} =  &-\frac{i}{\hbar}\left[ \hat{H},
 \hat{\rho}_t\right] \\ 
&-\lambda \frac{m^2}{m_0^2} \left(\frac{r_C}{\sqrt{\pi}\hbar}\right)^3 \int_0^t ds f(t-s) \int  d\bm{Q} e^{-\frac{\bm{Q}^2 r_C^2}{\hbar^2}} [e^{\frac{i}{\hbar}\bm{\hat{x}}\cdot \bm{Q}},[\hat{U}^\dagger(s-t)
e^{-\frac{i}{\hbar}\bm{\hat{x}}\cdot \bm{Q}} \hat{U}(s-t), \hat{\rho_t}]],
\end{split}
\end{equation}
where $f(t-s)$ is the correlation function and $\hat{U}(t)=e^{-\frac{i}{\hbar}\frac{\bm{\hat{p}}^2}{2m}t}$.

We now expand $\hat{U}(\tau)$ to first order: $\hat{U}(\tau)\approx1-\frac{i}{\hbar}\frac{\bm{\hat{p}}^{2}}{2m}\tau$,
which is justified since $\tau$ is limited by the correlation time
$\tau_{C}$ of the correlation function $f(s)$ through the time integral
in Eq.~\eqref{nw_equation_momentum}. We make the following assumption:
\begin{equation}
\frac{1}{\hbar}\frac{\bm{\hat{p}}^{2}}{2m}\tau_{C}\ll1.\label{eq:approx_markovian_condition}
\end{equation}
We can make a rough estimate for the maximum value of $\tau_{C}$
by replacing the operator with the expectation value in Eq.(\ref{eq:approx_markovian_condition}):
$\langle\bm{\hat{p}}^{2}/2m\rangle\tau_{C}/\hbar\ll1$, We consider
the temperature of the system to be $T\approx10^{2}-10^{3}K$. Thus
based on the equipartition theorem we replace $\langle\bm{\hat{p}}^{2}/2m\rangle$ by $k_{B}T$ which gives the condition $\tau_{C}<10^{-13}s$.
This gives us a corresponding minimum ultraviolet frequency cut-off
$\Omega\gg10^4\text{GHz}$ for the Fourier transform of the correlation
function.

Hence by performing the time integration we obtain from Eq.~\eqref{nw_equation_momentum}:
\begin{equation}
\label{nw_equation_app}
\frac{d \hat{\rho}_t}{dt} = L_{\text{\tiny CSL}}[\hat{\rho}] + L_{\text{\tiny correction}}[\hat{\rho}],
\end{equation}
where 
\begin{equation}
L_{\text{\tiny CSL}}[\hat{\rho}]=-\frac{i}{\hbar}\left[ \hat{H},
 \hat{\rho}_t\right] +\lambda \frac{m^2}{m_0^2}  \left( \left(\frac{r_C}{\sqrt{\pi}\hbar}\right)^3 
\int d \bm{Q} e^{-\frac{\bm{Q}^2r_C^2}{\hbar^2}}
e^{\frac{i}{\hbar} \bm{Q} \cdot \hat{\bm{x}}}\hat{\rho}(t)e^{-\frac{i}{\hbar} \bm{Q} \cdot \hat{\bm{x}}}- \hat{\rho}(t)\right) 
\end{equation}
is the white noise CSL evolution,
\begin{equation}\label{nw_correction}
\begin{split}
L_{\text{\tiny correction}}[\hat{\rho}]=
\frac{i \bar{\tau}}{2m\hbar} \lambda \frac{m^2}{m_0^2}   \left(\frac{r_C}{\sqrt{\pi}\hbar}\right)^3
\int  d\bm{Q}  e^{-\frac{\bm{Q}^2r_C^2}{\hbar^2}} \bm{Q} \cdot  &\left(\right. 
[e^{\frac{i}{\hbar} \bm{Q} \cdot \hat{\bm{x}}} \hat{\rho} e^{-\frac{i}{\hbar} \bm{Q} \cdot \hat{\bm{x}}}, \bm{\hat{p}}]
+e^{\frac{i}{\hbar} \bm{Q} \cdot \hat{\bm{x}}} [\hat{\rho} , \bm{\hat{p}}]e^{-\frac{i}{\hbar} \bm{Q} \cdot \hat{\bm{x}}}
\left. \right)
\end{split}
\end{equation}
is the first order correction due to the non white noise and 
\begin{equation}
\bar{\tau}=\int_0^t f(s)s ds.
\end{equation}

By performing a direct but tedious calculation, it can be shown that  Eq.~\eqref{nw_equation_app} is invariant under boost and thus fully Galilean invariant.

Let us now find the solution of Eq.~\eqref{nw_equation_app} by using the characteristic function approach~\cite{PhysRevA.82.042111}. We multiply Eq.~\eqref{nw_equation_app} by $e^{\frac{i}{\hbar} (\bm{\nu} \cdot \bm{\hat{x}} + \bm{\mu} \cdot \bm{\hat{p}})}$ and take the trace:
\begin{equation}\label{chi_eq}
\frac{\partial}{\partial t}\chi(\bm{\nu},\bm{\mu},t)= \frac{\bm{\nu}}{M} \cdot \partial_{\bm{\mu}}  \chi(\bm{\nu},\bm{\mu},t) 
+\lambda\left(\Phi(\bm{\nu},\bm{\mu}) -1\right),
\end{equation}
where
\begin{equation}
\Phi(\bm{\nu},\bm{\mu})=e^{\frac{-\mu^2}{4r_C^2}}(1-\frac{\bm{\mu} \cdot \bm{\nu}}{4mr_C^2}\bar{\tau} )
\end{equation}
and
\begin{equation}\label{chi_def}
\chi(\bm{\nu},\bm{\mu},t)= Tr[\hat{\rho}_t e^{\frac{i}{\hbar} (\bm{\nu} \cdot \bm{\hat{x}} + \bm{\mu} \cdot \bm{\hat{p}}}].
\end{equation}
The solution of the characteristic function in Eq.~\eqref{chi_eq} is given by:
\begin{equation}
\chi(\bm{\nu},\bm{\mu},t)= \chi^0(\bm{\nu},\bm{\mu},t)e^{-\lambda t
+ \int_0^t\Phi(\bm{\nu},\frac{\bm{\nu}\tau}{m}+\bm{\mu}) d\tau},
\end{equation}
where $\chi^0(\bm{\nu},\bm{\mu},t)$ is the solution of equation $\frac{\partial}{\partial t} \chi^0(\bm{\nu},\bm{\mu},t)= \frac{1}{m} \bm{\nu} \cdot \frac{\partial}{\partial  \bm{\mu}}\chi^0(\bm{\nu},\bm{\mu},t)$. 
The density matrix can be obtained from the characteristic function using the inversion formula:
\begin{equation}\label{inversion_formula}
\rho(\bm{x},\bm{x'},t)=\int \frac{d\bm{\nu}}{(2\pi\hbar)^3} 
e^{-\frac{i}{2\hbar} \bm{\nu} \cdot (\bm{x}+\bm{x'})} 
\chi(\bm{\nu},\bm{x}-\bm{x'},t).
\end{equation}
Hence the solution of the master equation \eqref{nw_equation_app} is given by:
\begin{equation}\label{Crho_grw_dp3}
\rho^{\text{\tiny cCSL}}(\bm{x},\bm{x'},t)=\frac{1}{(2\pi\hbar)^3} \int d\bm{\tilde{k}}  \int \bm{\tilde{y}} e^{-\frac{i}{\hbar}\bm{\tilde{k}} \cdot \bm{\tilde{y}}} 
F_{\text{\tiny cCSL}}(\bm{\tilde{k}},\bm{x}-\bm{x'},t) \rho^{QM}(\bm{x}+\bm{y},\bm{x'}+\bm{y},t),
\end{equation}
where
\begin{equation}\label{CFF_dp3}
F_{\text{\tiny cCSL}}(\bm{\tilde{k}},\bm{q},t)= F_{\text{\tiny CSL}}(\bm{\tilde{k}},\bm{q},t)
 \exp\Bigg[\frac{\lambda \bar{\tau}}{2}  \left(e^{-\frac{(\bm{q}-\frac{\bm{\tilde{k}}t}{m})^2}{4r_C^2}} - e^{-\frac{\bm{q}^2}{4r_C^2}}\right) \Bigg].
\end{equation}
For further details about the characteristic function approach see Ref.~\cite{PhysRevA.82.042111,PhysRevA.90.062135}. The interference pattern is given by Eq.~\eqref{p3_collapse} with the function $D$ replaced by $D_{\text{\tiny cCSL}}=D_{\text{\tiny CSL}}$ given in Eq.~\eqref{cfunc}. Although $F_{\text{\tiny CSL}}$ and $F_{\text{\tiny cCSL}}$ in general differ, i.e. CSL and cCSL have different free evolutions, we have the curious situation that the non Markovian effects in diffraction experiments cancel exactly, i.e. the CSL and cCSL interference patterns coincide.\\ \\

\noindent {\bf QMUPL: Quantum mechanics with universal position localization.} Here we are referring to the mass-proportional version of the QMUPL model~\cite{PhysRevA.40.1165,PhysRevA.42.5086}. The single-particle master equation in 3D is given by~\cite{PhysRevA.90.062105}:

\begin{equation}
\frac{d\hat{\rho}_t}{dt}=-\frac{i}{\hbar}\left[\hat{H},\hat{\rho}_{t}\right]-\frac{\eta}{2}\frac{m}{m_{0}}\left[\hat{\mathbf{x}},\left[\hat{\mathbf{x}},\hat{\rho}\right]\right]\label{eq:master}
\end{equation}
In the free particle case, the solution to this master equation can be obtained with the help of the characteristic function:
\begin{equation}
\rho^{\text{\tiny QMUPL}}(\bm{x},\bm{x}',t)=\frac{1}{(2\pi\hbar)^{3}}\int d\bm{\tilde{k}}\int \bm{\tilde{w}}e^{-\frac{i}{\hbar}\bm{\tilde{k}}\cdot\bm{\tilde{w}}}F_{\text{\tiny QMUPL}}(\bm{\tilde{k}},\bm{x}-\bm{x}',t)\rho^{\text{\tiny QM}}(\bm{x}+\bm{\tilde{w}},\bm{x}'+\bm{\tilde{w}},t),\label{eq:solution}
\end{equation}
where $\rho^{\text{\tiny QM}}(\bm{x},\bm{x}',t)$ denotes the usual
free quantum mechanical evolution ($\eta=0$) and

\begin{equation}
F_{\text{QMUPL}}(\bm{\tilde{k}},\bm{q},t)=\exp\left[\frac{\eta}{2}\frac{m}{m_{0}}\left[\mathbf{q^{2}}-\frac{\mathbf{q}\cdot\tilde{\mathbf{k}}t}{m}+\frac{\tilde{\mathbf{k}}^{2}}{m^{2}}\frac{t^{2}}{3}\right]\right].
\end{equation}
The interference pattern is given by Eq.~\eqref{p3_collapse} with
the function $D$ defined as follows: 
\begin{equation}
D_{\text{\tiny QMUPL}}(x_{2}-x_{2}')=\exp\left[-\frac{\eta}{3}\frac{m}{m_{0}}(t_{1}+t_{2})(x_{2}-x_{2}')^{2}\right].\label{cfunc-1}
\end{equation}
The function $D_{\text{QMUPL}}(q)$ completely encodes the modification
to the quantum mechanical interference pattern ($\eta=0$).

\clearpage
\section{Center-of-mass motion for a rigid object and the amplification mechanism}\label{sec_amp}
Matter-wave experiments use large molecules and create spatial superpositions of their center-of-mass motion. In this section, starting from the many-particle collapse dynamics, we will derive a closed equation for the center of mass, under the rigid-body approximation. We will show and quantify the amplification mechanism: the larger the system, the faster the collapse of the center-of-mass wave function.

We will start by considering the CSL model. Under siutable assumptions, discussed at the end of this section, the analysis  applies also to the dCSL,  cCSL with small correlation time, and to the DP model. We will discuss three approximations for the geometry of a planar molecule, namely Adler's formula~\cite{1751-8121-40-12-S03}, the homogeneous disk approximation and the 2D lattice structure approximation \cite{PhysRevLett.113.020405}. 

The $N$-particle CSL master equation reads:
\begin{equation}\label{many_particle_csl}
\frac{d }{d t}\hat{\rho}(t)= 
-\frac{i}{\hbar}\left[\hat{H},\hat{\rho}(t) \right]
+ \lambda  \frac{m^2}{m_0^2} \left(\frac{ r_C} {\sqrt{\pi} \hbar}\right)^3  \sum_{j,l}^{N} 
 \int d \bm{Q} e^{-\frac{\bm{Q}^2r_C^2}{\hbar^2}}
 \left(e^{\frac{i}{\hbar} \bm{Q} \cdot \hat{\bm{x}}_j}\hat{\rho}(t)e^{-\frac{i}{\hbar} \bm{Q} \cdot \hat{\bm{x}}_l}- \hat{\rho}(t)\right),
\end{equation}
where $m$ is the mass of a single particle and $\bm{\hat{x}}_i$ is the position operator of particle $i$. By performing a trace over the relative coordinates, we obtain the master equation for the reduced density matrix $\hat{\rho}_{\text{\tiny CM}}(t)$ describing the center-of-mass motion:
\begin{equation}\label{cm_csl}
\begin{split}
\frac{d }{d t}\hat{\rho}_{\text{\tiny CM}}(t) &= 
-\frac{i}{\hbar}\left[\hat{H},\hat{\rho}_{\text{\tiny CM}}(t) \right]\\
&+\lambda \left(\frac{ r_C} {\sqrt{\pi} \hbar}\right)^3 \frac{m^2}{m_0^2} 
\int d \bm{Q} R(\bm{Q}) e^{-\frac{\bm{Q}^2r_C^2}{\hbar^2}}
(e^{\frac{i}{\hbar} \bm{Q} \cdot \hat{\bm{X}}}\hat{\rho}_{\text{\tiny CM}}(t)e^{-\frac{i}{\hbar} \bm{Q} \cdot \hat{\bm{X}}}- \hat{\rho}_{\text{\tiny CM}}(t)),
\end{split}
\end{equation}
where $\bm{\hat{X}}=\sum_{i=1}^{N}\bm{\hat{x}}_i/N$ is the center of mass position operator and
\begin{equation}
R(\bm{Q})=\int  d\bm{r}_1...d\bm{r}_N \sum_{j=1,l=1}^{N} e^{\frac{i}{\hbar} \bm{Q} \cdot (\bm{r}_j-\bm{r}_l)}
\end{equation}
encodes  the distribution of atoms in space around the center of mass. By considering a rigid body and neglecting rotations around the center of mass, we can remove the integrations over the relative coordinates~\cite{PhysRevA.42.78}:
\begin{equation}\label{R_function}
R(\bm{Q})= \sum_{j=1,l=1}^{N} e^{\frac{i}{\hbar} \bm{Q} \cdot (\bm{r}_j-\bm{r}_l)}.
\end{equation}
The next step  is to replace $R(\bm{Q})$ with a function independent of the position of the particles, so that Eq.~\eqref{cm_csl} reduces to a single-particle master equation like Eq.~\eqref{csl_equation}, with $\lambda$  replaced by an enhanced factor  $\Lambda$, which depends on the total number of particle and their geometrical distribution. Hence we want to show that under suitable approximations: 
\begin{align}
\lambda \frac{m^2}{m_0^2} \int d \bm{Q} R(\bm{Q}) e^{-\frac{\bm{Q}^2r_C^2}{\hbar^2}} 
e^{\frac{i}{\hbar} \bm{Q} \cdot \hat{\bm{X}}}\hat{\rho}_{CM}(t)e^{-\frac{i}{\hbar} \bm{Q} \cdot \hat{\bm{X}}}& &\xrightarrow{}& \;\;\;\;\;\Lambda  \int d \bm{Q} e^{-\frac{\bm{Q}^2r_C^2}{\hbar^2}}
e^{\frac{i}{\hbar} \bm{Q} \cdot \hat{\bm{X}}}\hat{\rho}_{CM}(t)e^{-\frac{i}{\hbar} \bm{Q} \cdot \hat{\bm{X}}},\label{withxrescaling}\\
-\lambda \frac{m^2}{m_0^2}  \left(\frac{ r_C} {\sqrt{\pi} \hbar}\right)^3  \int d \bm{Q} R(\bm{Q}) e^{-\frac{\bm{Q}^2r_C^2}{\hbar^2}}
\hat{\rho}_{CM}(t)& &\xrightarrow{}&\;\;\;\;\;-\Lambda  \hat{\rho}_{CM}(t)\label{withoutxrescaling}.
\end{align}
We now review the three possible methods of approximation mentioned above. \\ \\

\begin{figure}[!htb]
\begin{center}
\def\svgwidth{260pt}
\includegraphics[width=0.35\textwidth]{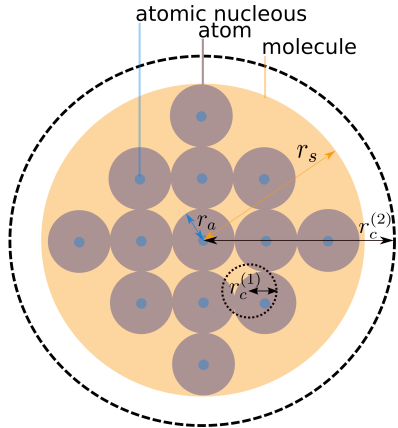}
\caption{Macro-molecule thin disk approximation with uniformly distributed atoms. The blue circles represent atomic nuclei, the purple circles the atoms of radius $r_a$ and the orange circle of radius $r_s$ denotes the area spanned by the molecule. We assume for simplicity, that the purple circles denoting atoms completely fill the orange circle denoting the molecule, so that empty spaces can be neglected. When $r_C>r_s$, the whole molecule is contained within a circle of radius $r_C$, and the quadratic scaling law  applies (e.g.  $r_C=r_C^{(2)}$) . When $r_C<r_a$, only one nucleus  is contained within a circle of radius $r_C$, and the linear scaling law  applies (e.g. for $r_C=r_C^{(1)}$). When $r_a<r_C<r_s$, we interpolate the two limiting cases with the scaling law \eqref{mass2}. }
\label{molecular_model}
\end{center}
\end{figure}

\noindent {\bf Adler's formula}.
Consider first the situation when the molecule is enclosed in a radius $r_s \ll r_C$ (see Fig.~\ref{molecular_model}). According to Eq.~\eqref{cm_csl} the weight $
 \exp(-{\bm{Q}^2r_C^2/\hbar^2})$ selects those values of $|\bm{Q}|$ such that  $|\bm{Q}| < \hbar/r_C$.  Hence we have that $|\frac{1}{\hbar} \bm{Q} \cdot (\bm{r_j}-\bm{r_l})| < |\frac{(\bm{r_j}-\bm{r_l})}{r_C}| \approx 0$, and we can write:
\begin{equation} \label{eq:dhggg}
R(\bm{Q}) \approx \sum_{j=1,l=1}^{N} 1= N^2.
\end{equation}

On the opposite side, let us consider the situation when the distance between nearest neighbour atoms $r_a$ is much bigger than $r_C$, .i.e. $r_C\ll r_a$. We group the terms $e^{-\bm{Q}^2r_C^2/\hbar^2} 
 (e^{\frac{i}{\hbar} \bm{Q} \cdot (\bm{r}_j-\bm{r}_l)}+ e^{-\frac{i}{\hbar} \bm{Q} \cdot (\bm{r}_j-\bm{r}_l)})$, which can be rewritten as:
$ 2 e^{-\bm{Q}^2r_C^2/\hbar^2} \cos(\bm{Q} \cdot (\bm{r}_j-\bm{r}_l)/\hbar)$. Let us rewrite: 
$\bm{Q} \cdot (\bm{r}_j-\bm{r}_l) =|\bm{Q}| |\bm{r}_j-\bm{r}_l| \cos(\theta)$. Except for the cases when $\cos(\theta) \approx 0$, if $j \neq l$ the condition  $r_C\ll r_a$ implies that the oscillations of $\cos(|\bm{Q}| |\bm{r}_j-\bm{r}_l| cos(\theta) /\hbar )$ make the $\bm{Q}$ integrals negligible. Therefore, the dominant contribution in Eq.~\eqref{R_function} comes from $j=l$ terms, and we can write:
\begin{equation}\label{rq_eq}
R(\bm{Q}) \approx \sum_{j=1}^{N} 1= N.
\end{equation}
 
The conclusion is that, when $N$ particles in the system are distant less than $r_C$,  we have a quadratic scaling ($\sim N^2$) of $\Lambda$ for the center of mass motion. On the other hand, when the mutual distance between the $N$ particles is larger than $r_C$, then $\Lambda$ for the center of mass motion increases linearly with $N$. 

We also need to consider the intermediate case, where a more careful analysis is needed. In this situation, the behaviour is expected to interpolate between the linear and quadratic scalings. We model the macro-molecules used in the experiments by atoms uniformly distributed over a thin disk, as depicted in Fig.~\ref{molecular_model}. We neglect the electrons, as their mass is small compared to the nucleon mass and we describe the atomic nuclei as single particles of average mass $m_a=\frac{m}{n_a}$ (average atomic mass), where $n_a$ is the total number of atoms. We limit the discussion to values of $r_C$ larger than the nucleon size $\sim 10^{-15}\text{m}$.
The mean area covered by a single atom is $\pi r_a^2$, where we take the mean atomic radius to be $r_a= 10^{-10}m$. The number of atoms contained within a circle of radius $r_C$ is:
\begin{equation}
  n(r_C)=\begin{cases}
    1, & \text{if $ r_C< r_a$}.\\
    \frac{\pi r_C^2}{\pi r_a^2}, & \text{if $r_a \leq r_C\leq r_s$}.\\
    n_a & \text{if $r_s < r_C$}.
  \end{cases}
\end{equation}
These will contribute quadratically to the collapse rate.  The molecule can be covered by $n_a/n(r_C)$ circles of radius $r_C$ and atoms belonging to different circles contribute linearly to the collapse rate. Thus we model the collapse rate for the center of mass of the molecule according to the formula:
\begin{equation}\label{mass2}
\Lambda= \frac{n_a}{n(r_C)} \left(\frac{m_a n(r_C)}{m_0} \right)^2\lambda.
\end{equation}
This is the formula we will use in following sections. We will describe the center of mass motion as that of a single particle via Eq.~\eqref{rho_3d}, and in all formulas derived starting from it, $\lambda \left(\frac{m}{m_0}\right)^2$ is replaced by $\Lambda$. Of course, in the limiting case when the molecular radius $r_s$ is smaller then $r_C$, the above scaling reduces to the purely quadratic scaling law, while when the atomic radius $r_a$ is larger than $r_C$ it reduces to the purely linear scaling law. 
We now discuss two further approximation schemes which will confirm the validity of Eq.~\eqref{mass2}.
\\ \\

\noindent {\bf Homogeneous thin disk approximation.}
As a different way to tackle the problem, let us consider the molecule as a thin \textit{homogeneous} disk of radius $r_s$ and thickness $d$. In this continuous limit, we can approximate:
\begin{equation}
\sum_{j=1}^{N} e^{\frac{i}{\hbar} \bm{Q} \cdot \bm{r_j}}  \quad \longrightarrow \quad  \int d\bm{x} \rho_{\text{\tiny rel}}(\bm{x}) e^{\frac{i}{\hbar} \bm{Q} \cdot \bm{x}} = \tilde{\rho}_{\text{\tiny rel}}(\bm{Q}),
\end{equation}
where $\rho_{\text{\tiny rel}}(\bm{x}) $ is the matter distribution around the center of mass, and $\tilde{\rho}_{\text{\tiny rel}}(\bm{Q})$ its Fourier transform. Then Eq.~\eqref{R_function} reduces to:
\begin{equation}\label{homogeneousR}
R(\bm{Q})= |\tilde{\rho}_{\text{\tiny rel}}(\bm{Q})|^2.
\end{equation}
In particular, by labelling the axis of rotational symmetry of the disk as the $z$ axis and its orthogonal plane ($x$-$y$ plane) with the label $o$, we find that 
\begin{equation}
\tilde{\rho}_{\text{\tiny rel}}(\bm{Q})=\frac{2\hbar}{Q_o R} J_1(\frac{Q_o R}{\hbar}) \sinc(\frac{Q_z d}{2 \hbar}),
\end{equation} 
where $Q_z$,$Q_o$ are the $z$ axis and the $x$-$y$ plane components of $\bm{Q}$, respectively and $J_1$ denotes the Bessel function of the first kind. We now insert $\tilde{\rho}_{\text{\tiny rel}}(\bm{Q})$ into Eqs.~\eqref{withxrescaling} and \eqref{withoutxrescaling} and take the limit $d \rightarrow 0$ (very thin disk approximation). To perform the approximation in Eq.~\eqref{withxrescaling} and Eq.~\eqref{withoutxrescaling}, we work in the position basis, i.e. we apply $\langle x, y, z |$, $|x', y', z'\rangle$ from the left and right, respectively. In addition, we assume that the superposition is on distances much greater than the size of the system, i.e. $\text{$\Delta $x} = x-x'$ is either $|\text{$\Delta $x}|\gg r_s$ or $\text{$\Delta $x}=0$ and similarly for the $y$ axis. It is then easy to obtain the rescaling of the parameter $\lambda$:
\begin{equation}
\Lambda=\frac{4 \lambda m^2 r_C^2 \left(1-e^{-\frac{r_s^2}{4 r_C^2}}\right)}{m_0^2 r_s^2}.
\end{equation}\\ 

\noindent {\bf 2D Lattice disk}.
As a different approximation, we consider a 2-D lattice, as depicted  in Fig.~\ref{molecular_model}, of point-like nuclei (small blue circles) forming a thin disk  of radius $r_s$ (orange circle).  The axis of rotational symmetry of the disk is $z$, the nuclei sit on the $x$-$y$ plane and their poisition is denoted as $(n_x,n_y)$. The index $n_x$ runs from $n_{min}=-\lfloor \frac{r_s}{a} \rfloor$ to $n_{max}=\lfloor \frac{r_s}{a} \rfloor$, where we take $a=10^{-10}\text{m}$ to be the lattice constant and $\lfloor . \rfloor $ indicates the floor rounded value. Hence the $n_y$ index runs from $-\left\lfloor\sqrt{\frac{r_s^2}{a^2}-n^2}\right\rfloor$ to $\left\lfloor \sqrt{\frac{r_s^2}{a^2}-n^2}\right\rfloor$ in accordance with the circular shape of the molecule $n_x^2 +n_y^2 \leq \left(\frac{r_s}{r_a}\right)^2$.  In other words, we consider the following $R(\bm{Q})$ function \eqref{R_function}:
\begin{equation}\label{R_function_sums}
R(\bm{Q})=
\sum _{\substack{n_x^2+n_y^2 \leq n_{max}^2 \\
n_x^{\prime 2}+n_y^{\prime 2} \leq n_{max}^2}} 
e^{ \frac{i}{\hbar}a(n_x-n'_x) Q_x  +  \frac{i}{\hbar}a(n_y-n'_y) Q_y } 
\end{equation}
where the primed and undprimed variables label the first and second sum, respectively.

Let us first deal with the rescaling in Eq.~\eqref{withoutxrescaling}. We perform the $d\bm{Q}$ integration and we get the rescaled parameter $\Lambda$:
\begin{equation}\label{no_x_term}
\Lambda=\lambda 
\sum _{\substack{n_x^2+n_y^2 \leq n_{max}^2 \\
n_x^{\prime 2}+n_y^{\prime 2} \leq n_{max}^2 }}
\exp \left(
-\frac{a^2 \left(\Delta n_x\right)^2}{4 r_C^2}
-\frac{a^2 \left(\Delta n_y\right)^2}{4 r_C^2}
\right).
\end{equation}
where $\Delta n_x=n_x-n'_x$ and $\Delta n_y=n_y-n'_y$.

Next, we consider the rescaling in Eq.~\eqref{withxrescaling}. To ease the analysis, we work in the position basis, i.e. we apply $\langle x, y, z |$, $|x', y', z' \rangle$ from the left and right, respectively. We consider a single term in Eq.~\eqref{R_function_sums} and perform the $d\bm{Q}$ integration in Eq.~\eqref{withxrescaling}, we get:
\begin{equation}\label{single_term_with_x}
\lambda \exp \left(-\frac{(a \Delta n_x+\Delta x)^2}{4 r_C^2}-\frac{(a \Delta n_y+\Delta y)^2}{4 r_C^2}-\frac{\Delta z ^2}{4 r_C^2}\right),
\end{equation}
where $\Delta x,\Delta y,\Delta z$ are $x-x',y-y',z-z'$ respectively. Let us again assume that the superposition varies on distances much greater than the size of the system, i.e. $\text{$\Delta $x}$ is either $|\text{$\Delta $x}| \gg a \Delta n$ or $\text{$\Delta $x}=0$. Hence we can approximate  $\left(a \text{$\Delta $n}+\text{$\Delta $x}\right){}^2  \approx \left(a \text{$\Delta $n} \right){}^2+ \left(\text{$\Delta $x}\right){}^2$. A similar argument can be carried also for the $y$ axis variables. Thus, combining Eq.~\eqref{no_x_term} and Eq.~\eqref{single_term_with_x} we finally obtain:
\begin{equation}\label{finally_x_term}
\Lambda \exp \left(-\frac{ \text{$\Delta $x}{}^2+\text{$\Delta $y}{}^2+\text{$\Delta $z}{}^2}{4 r_C^2}  \right) -\Lambda,
\end{equation}
which implies that the center of mass density matrix satisfies the one particle CSL master equation with the rescaled parameter $\Lambda$.
\label{equation}
\\ \\

\begin{figure}[!htb]
\begin{center}
\includegraphics[width=1.0\textwidth]{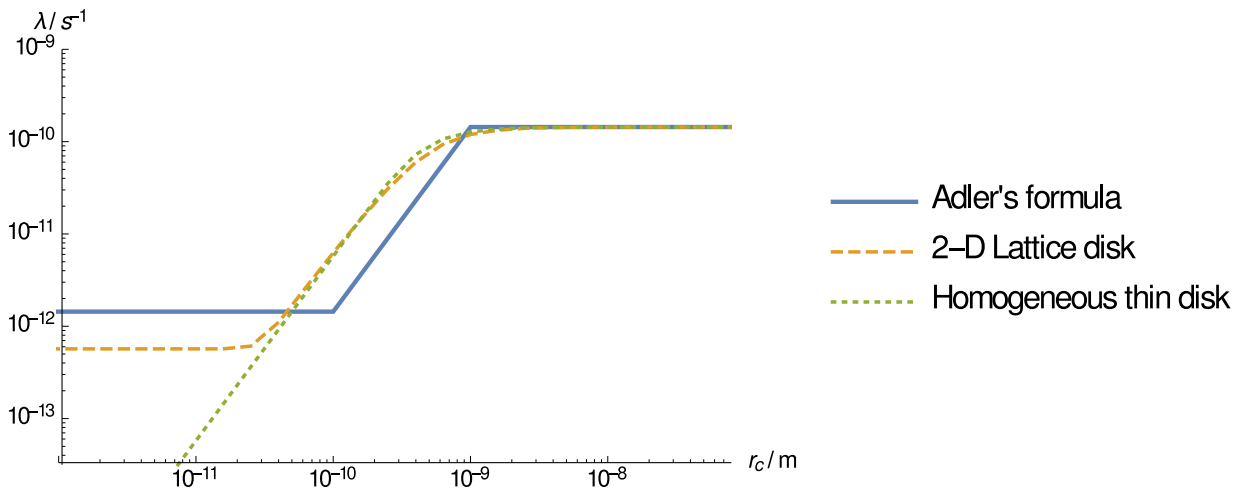}
\caption{Amplification of the parameter $\Lambda$ as a function of $r_C$ for three different approximations: Adler's formula (solid blue line), homogenous disk approximation (dotted green line), 2D lattice approximation 
(dashed orange line). We see, that Adler's formula agrees very well with the more sophisticated 2D lattice model approximation, while the homogenous disk approximation breaks down at distances below the atomic radius ($r_a=10^{-10}\text{m}$). The plot is obtained for $N=100$ atoms  with atomic (nuclei) mass $12 m_0 =12 \text{amu}$.}
\label{amplification_mechanisms}
\end{center}
\end{figure}

\noindent {\bf Comparison and other collapse models}.
The three approximations discussed here above are compared in Fig.~\ref{amplification_mechanisms}. 
In particular, we see that Adler's heuristic formula is in good agreement with the 2D lattice model amplification mechanism.  We also see that the homogeneous thin disk approximation begins to break down for $r_C$ values smaller than the atomic radius $r_a$ as one would expect.

We also stress the key assumption used in the derivation of the amplification mechanism: $r_s \ll r_{sup}$, where $r_s$ is the size of the system (e.g. molecular radius) and $r_{sup}$ is the size of the macroscopic superposition. Only using this assumption, we were able to effectively describe the center of mass motion master equation \eqref{cm_csl} by the single particle master equation \eqref{csl_equation} with the rescaled parameter $\Lambda$. When $r_s \gtrsim  r_{sup}$ we have a weaker suppression of macroscopic superpositions. 

The dCSL and cCSL models (with small correlation time) many particle master equations have a similar structure as that of the CSL model. Hence, as in part argued in Refs.~\cite{Smirne:2014paa},\cite{1751-8121-40-50-012}, the amplification mechanism is analogous to the CSL amplification mechanism. For the dCSL model, one has to also consider the parameter $k_T$, which limits the validity of the approximations.

The previous analysis is also applicable to the  DP model, as it can be easily shown by considering the many particle DP master equation. As previously stated, in the DP model, $R_0$ can be identified with $r_C$ and $\lambda$ can be identified with $\frac{G m_0^2}{\hbar \sqrt{\pi} R_0}$. As in the CSL model $\lambda$ rescales to $\Lambda(r_C)$, in the DP model  $\frac{G m_0^2}{\hbar \sqrt{\pi} R_0}$ rescales to  $\lambda_{DP}=\frac{G m_0^2}{\hbar \sqrt{\pi} R_0} \frac{\Lambda(R_0)}{\lambda}$.  

\begin{figure}[!htb]
\begin{center}\label{lambda_scaling}
\includegraphics[width=0.5\textwidth]{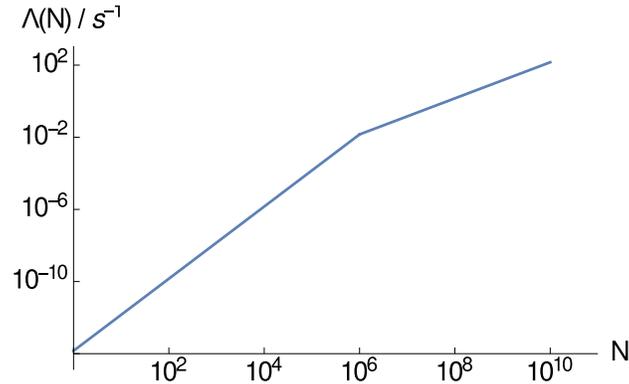}
\caption{The plot shows the amplification of the effective collapse rate $\Lambda$ according to Eq.~\eqref{mass2} for the thin disk model described in the text. The plot is obtained with $\lambda=10^{-16}\text{s}^{-1}$, $r_C=10^{-7}\text{m}$, atomic radius $r_a=10\text{m}$ and atomic mass $m_a=12m_0=12 \text{amu}$. We notice that at $N=10^6$ the amplification mechanism changes behavior as the total size of the system $r_s$  becomes equal to $r_C$.}
\end{center}
\end{figure}

In Fig.~\eqref{lambda_scaling} we show how $\Lambda$ varies as a function of the total number of atoms $N$  according to Eq.~\eqref{mass2}, where the atoms form a thin disk lattice structure, as described in Fig.~\ref{molecular_model}, 

The GRW and dGRW models have a simple linear scaling of $\Lambda$ with the mass of the system by construction. \\ \\

\noindent {\bf Localization requirement of macroscopic objects}.
In the next section we will derive the upper bounds on the collapse parameters, with reference to the KDTL experiment, but there are also lower bounds, as the collapse cannot be too weak, otherwise the model looses its usefulness. The basic requirement for any collapse model is the rapid suppression of  macroscopic superpositions. We make the following reasonable, although arbitrary, minimal request: a macroscopic superposition of an object, visible by the naked eye (with spatial resolution $r$), should decay within a short time, set by the temporal resolution $t$ of the eye. This implies for example that a macroscopic superposition for a single-layered Graphene disk of radius $r$ localizes with an effective rate $t^{-1}$. 

The quantitative analysis is carried out in the following way. We neglect the free quantum mechanical evolution, while retaining the modification due to the collapse dynamics, i.e we neglect $\hat{H} = \bm{\hat{p}}^2/2m$. This is a reasonable assumption since the free quantum mechanical evolution is negligible for macroscopic objects on the time scale during which the wave function localizes. We have solved the resulting dynamics for each of the collapse models using the characteristic function approach~\cite{PhysRevA.82.042111}. For each of the considered collapse models the corresponding characteristic function equation is given by: 
\begin{equation}\label{chi_eq_localization}
\frac{\partial}{\partial t}\chi(\bm{\nu},\bm{\mu},t)= \Lambda\left(\Phi(\bm{\nu},\bm{\mu}) -1\right),
\end{equation}
where $\Phi$ depends on the model. We can easily obtain the solution to this equation:
\begin{equation}\label{chi_eq_solution}
\chi(\bm{\nu},\bm{\mu},t)= \chi(\bm{\nu},\bm{\mu},0) 
\exp \left(-\Lambda t(1- \Phi(\bm{\nu},\bm{\mu}))\right),
\end{equation}
Using the inversion formula given by Eq.~\eqref{inversion_formula} we obtain the corresponding density matrix: 
\begin{equation}\label{Crho_grw_dp3_localization}
\rho(\bm{x},\bm{x'},t)=\frac{1}{(2\pi\hbar)^3} \int d\bm{\tilde{k}}  \int \bm{\tilde{w}} e^{-\frac{i}{\hbar}\bm{\tilde{k}} \cdot \bm{\tilde{y}}} \exp \left( -\Lambda t(1- \Phi(\bm{\tilde{k}},\bm{x}-\bm{x'})) \right)
 \rho(\bm{x}+\bm{w},\bm{x'}+\bm{w},0),
\end{equation}
where $\rho(\bm{x}+\bm{w},\bm{x'}+\bm{w},0)$ is the initial density matrix.

Formally, we can also obtain the solution of the collapse dynamics (without the free quantum mechanical term)  from the full solution (with the free quantum mechanical term) by taking the limit  $m \rightarrow \infty$ in the expressions originating from the free quantum mechanical evolution, while keeping finite $m$ in the other expressions. 

We now list the solutions for the considered collapse models using the notation of  section \ref{socm}. For the CSL we obtain:
\begin{equation}\label{localization_csl}
\rho_{\text{\tiny CSL}}(\bm{x},\bm{x}',t)=\rho(\bm{x},\bm{x}',0) \exp \left(-\Lambda t \left(1-e^{-\frac{(\bm{x}-\bm{x})^2}{4r_C^2}}\right) \right).
\end{equation} 
The same formula applies also for the cCSL model with small correlation times $\tau_C$.

For the QMUPL we obtain:
\begin{equation}
\rho_{\text{\tiny QMUPL}}(\bm{x},\bm{x}',t)=\rho(\bm{x},\bm{x}',0)\exp\left(-\eta\frac{m}{m_{0}}t(\bm{x}-\bm{x}')^{2}\right).\label{eq:solution2}
\end{equation}

For the DP we obtain:
\begin{equation}
\rho_{\text{\tiny DP}}(\bm{x},\bm{x}',t)=\rho(\bm{x},\bm{x}',0) \exp 
\left(-\frac{t}{\hbar} \left(U(\bm{x}-\bm{x}')-U(0)\right) \right).
\end{equation} 

For the dCSL we obtain:
\begin{equation}\label{rho_grw_dp3_localization}
\begin{split}
&\rho_{\text{\tiny dCSL}}(\bm{x},\bm{x'},t)= \frac{1}{(2\pi\hbar)^3}\int d\bm{\tilde{w}} \rho(\bm{x}+\bm{\tilde{w}},\bm{x'}+\bm{\tilde{w}},0)  \\
& \int d\bm{\tilde{k}} e^{-\frac{i}{\hbar}\bm{\tilde{k}} \cdot \bm{\tilde{w}}} \exp \left(\right.-\Lambda t(1- e^{-\frac{\bm{\tilde{k}}^2 r_C^2 k_T^2}{\hbar^2}}
 e^{-\frac{(\bm{x}-\bm{x'})^2}{4r_C^2 (1+k_T)^2}}  e^{\frac{i}{\hbar}\frac{2 k_T m \bm{u}}{(1+k_T)} \cdot (\bm{x}-\bm{x}')} ) \left.\right),
\end{split}
\end{equation}
where in the limit $k_T \rightarrow 0$ we obtain the CSL solution given by Eq.~\eqref{localization_csl}. While for the CSL, cCSL and DP models we were able to perform the $\bm{\tilde{k}}$ and $\bm{\tilde{w}}$ integrations, for the dCSL the two integrations in general cannot be performed analytically. Hence for the dCSL we do not have in general a simple exponential decay of the off-diagonal elements. However, we can still investigate the dCSL decay of the off-diagonal elements by considering a particular initial state and performing a numerical simulation. In particular, we have considered a superposition state of two Gaussians centered at points $(r/2,0,0)$ and  $(-r/2,0,0)$:
\begin{equation}
\psi(\bm{x},0)= \left(\exp[-\frac{(x-r/2)^2}{4\sigma^2}]+\exp[-\frac{(x+r/2)^2}{4\sigma^2}]\right)\exp[-\frac{y^2}{4\sigma^2}]\exp[-\frac{z^2}{4\sigma^2}]
\end{equation}
with $r$ the spatial resolution of the eye and $\sigma=10^{-5}\text{m}$.

For the considered collapse models we can thus write the localization requirement for macroscopic objects as an inequality:
\begin{equation}
\biggr\lvert \frac{\rho\left(\bm{x},\bm{x}',t\right)}{\rho\left(\bm{x},\bm{x}',0\right)} \biggr\rvert  < \exp(-1),
\end{equation}
where we set $\bm{x}=(r/2,0,0)$, $\bm{x}'=(-r/2,0,0)$ and $t$ and $r$ are the eye temporal and spatial resolutions, respectively. The constant $\exp(-1) \sim 0.37$ is chosen arbitrarily, reflecting that for most collapse models the decay of the off-diagonal elements is exponential. This inequality will be used to obtain bounds on collapse parameters.

The same analysis applies also for the GRW and dGRW models, the only difference being the amplification mechanism discussed before.

\clearpage
\section{Experimental data analysis}\label{desec} 
We are now ready to apply the above results to the experiments \cite{MSCH} and \cite{C3CP51500A}. For concreteness, we illustrate the procedure with the CSL model. The same procedure is applicable for each collapse model described in Sec.~\ref{socm}.

In these experiments one has a source of molecules that have different velocities $v$ along the optical axis $z$.  Hence the real far-field interference pattern is given by:
\begin{equation}\label{p3_far_final}
\int_0^{+\infty} p_f(v) p(x;v) dv,
\end{equation}
where $p$ is given by Eq.~\eqref{p3_far} and $p_f(v)$ is the macromolecule velocity profile.
Similarly the real near-field interference pattern is given by
\begin{equation}\label{p3_near_final}
\int_0^{+\infty} p_n(v) S(x_{3s};v) dv,
\end{equation}
where $S$ is given by Eq.~\eqref{p3_near} and $p_n(v)$ is the macromolecule velocity profile.

To make a quantitative comparison with experimental data, we consider a grid of  pairs $(\lambda,r_C)$ and for each pair we perform a $\chi^2$ minimization procedure for the predicted CSL pattern according to Eqs.~\eqref{p3_far_final} and \eqref{p3_near_final}. In this way, we obtain a parameter diagram with an exclusion zone of pairs $(\lambda,r_C)$ that are incompatible with experimental data. 

A note of caution is at order. We have initially attempted to fit the experimental data by adopting the Poisson experimental error $\sqrt{I}$ for each value $I$ recorded by the detector since error bars were not reported in the papers. With this choice we were unable to obtain reasonable values of $\chi^2$ even for the standard quantum mechanical predictions. This is probably due, at least in part, to the approximations in the theoretical modeling and to unknown sources of error in the experiment. In order to circumvent this problem and to obtain reasonable values of $\chi^2$, we used an enlarged Poisson experimental error $a \sqrt{I}$, where $a$ is a constant. In order for the standard quantum mechanical fits to have reasonable $\chi^2$ values, we took $a=4.5$ for both experiments, but different values of a (within the same magnitude) do not change the final result. \\ \\   

\noindent {\bf Far-field}.
We first analyze the interference experiment with Phthalocyanine $C_{32}H_{18}N_{8}$ molecules reported in \cite{Real}, with the date taken from Ref.~\cite{MSCH}. The experimental setup is shown in Fig.~\ref{fig:experimental_setups}. The velocity profile was estimated according to Ref.~\cite{MSCH}. One has to be careful in considering the van der Walls forces between the molecules and the grating. This is modelled by considering an effective slit width smaller than the real one as described in \cite{Real}. The effective value is  $l_{eff}=43nm$. The  finite spatial resolution of the detector $4 \mu m$ was also taken into account.

\begin{figure}[!htb]
\centering
\begin{minipage}[t]{0.5\textwidth}
	\includegraphics[width=1.0\linewidth]{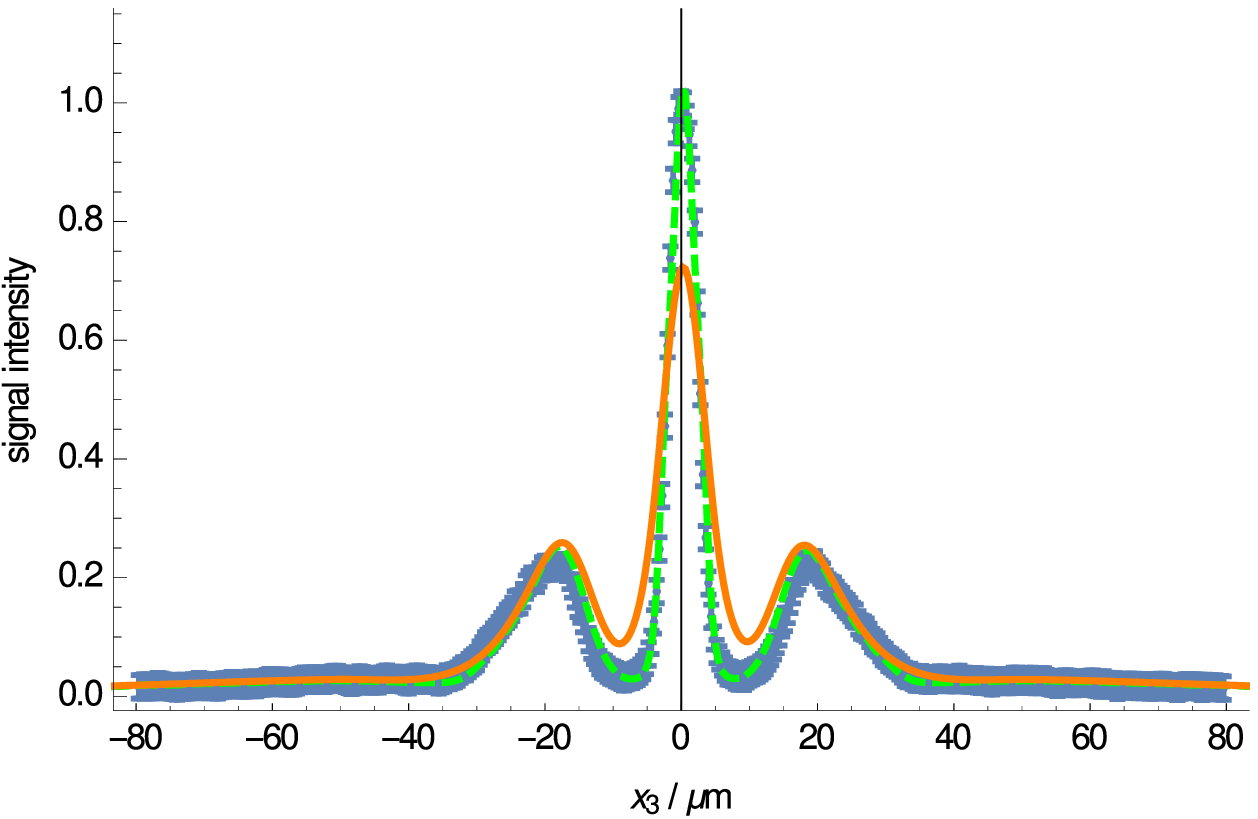}
\end{minipage}%
\begin{minipage}[t]{0.5\textwidth}
	\centering
  \includegraphics[width=1.0\linewidth]{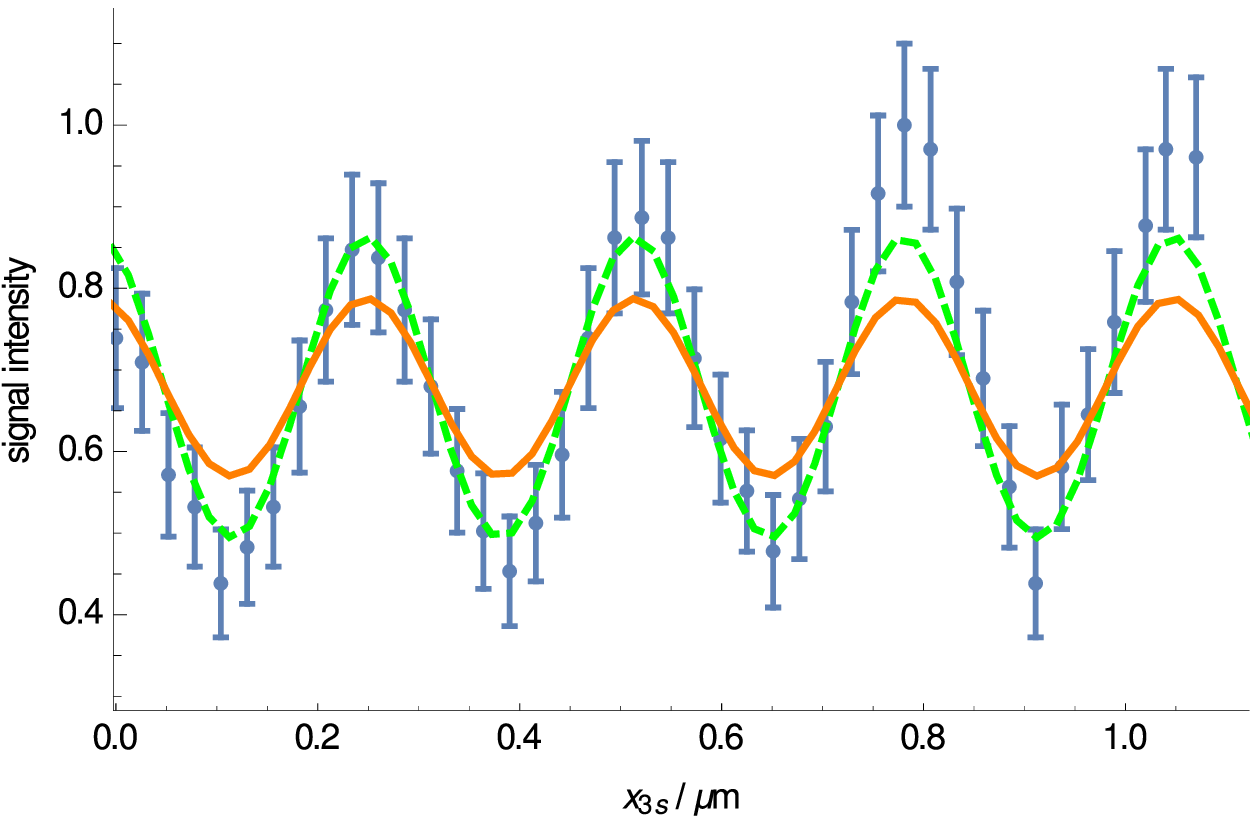}
\end{minipage}
\caption{Left: Far-field experiment \cite{MSCH}: $\lambda \approx 3.8 \cdot 10^{-3} \text{s}^{-1}$ and $r_C=10^{-7} \text{m}$. Right: KDTL near-field experiment \cite{C3CP51500A}:  $\lambda \approx 0.98 \cdot 10^{-5} \text{s}^{-1}$, $r_C=10^{-7} \text{m}$, laser power $P_{laser}=1\text{W}$. The orange dashed line represents the quantum mechanical fit, the solid orange line represents the CSL fit for an arbitrarily chosen (large) parameter $\lambda$ and the conventional $r_C$ value and the blue points and blue error bars represent the experimental data. The $y$ axis values are rescaled such that the maximum value is equal to unity.}
\label{fig:interference_figures}
\end{figure}

\begin{figure}[!htb]
\begin{center}
\includegraphics[width=0.65\textwidth]{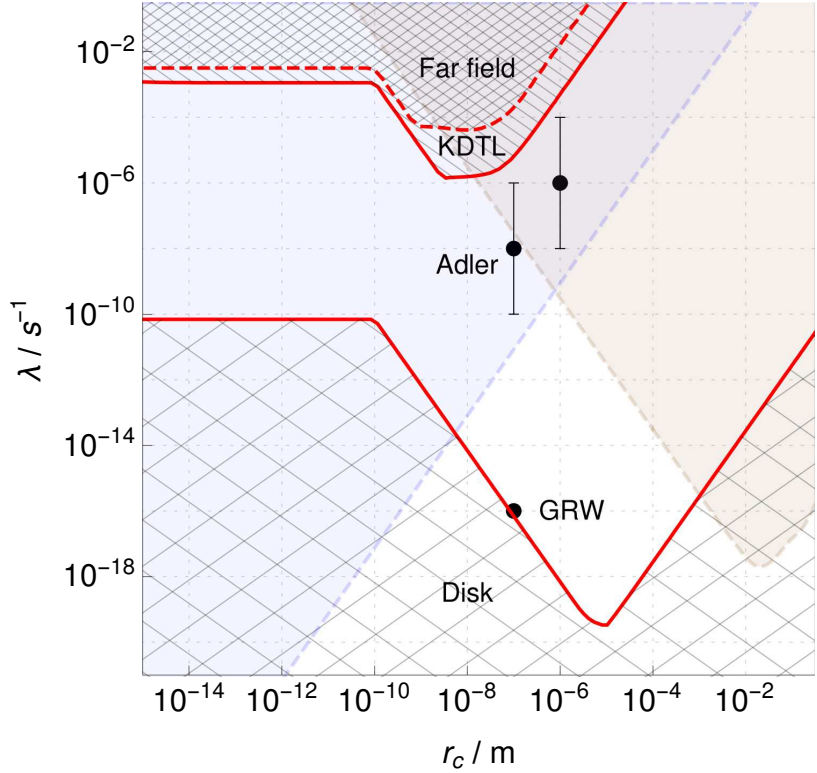}
\caption{Parameter diagram for the CSL, dCSL and cCSL models. The shaded exclusion zone delimited by the red lines applies to the CSL model as well as to the cCSL ($\Omega \gg 10^{13}Hz$) and dCSL models within specific limits of validity (see Figs.~\ref{limitsT} and \ref{limitsTB}). The lower exclusion zone originates from the requirement that a single-layered Graphene disk of radius $r=0.01mm$ is localized within $t=10ms$ (see Sec.~\ref{sec_amp}). The top zone delimited by the red solid (dashed) line  is excluded by the KDTL macromolecule interferometry experiment \cite{C3CP51500A} (far-field macromolecule interferometry experiment~\cite{Real} with the data taken from~\cite{MSCH}). For comparison we have included the upper bounds from X-ray experiments \cite{Curceanu}, valid for the CSL model and the cCSL model with frequency cutoff $\Omega \gg 10^{18}Hz$, and  the upper bounds from LISA Pathfinder~\cite{carlesso2016experimental}, valid for the CSL model only, so far: the exclusion zones are denoted by light blue and light orange colors, respectively. We have also included for reference, the GRW~\cite{PhysRevD.34.470} values $(\lambda=10^{-16}\text{s}^{-1},r_{C}=10^{-7}\text{m})$ and the values proposed by Adler \cite{1751-8121-40-12-S03}:  $(\lambda=10^{-8\pm2} \text{s}^{-1},r_{C}=10^{-7}\text{m})$ and $(\lambda=10^{-6\pm2}\text{s}^{-1},r_{C}=10^{-6}{m})$.}
\label{CSL_parameter_diagram}
\end{center}
\end{figure}

As an example, in Fig.~\ref{fig:interference_figures} we plot a comparison between the experimental interference pattern, the quantum mechanical fit and the CSL fit, for some arbitrarily chosen pair of parameters $\lambda, r_C$.
More importantly, we repeated the simulation for different pairs of parameters $\lambda, r_C$ as described before, obtaining the CSL parameter diagram shown in Fig.~\ref{CSL_parameter_diagram}.\\

\noindent {\bf Near-field  KDTL}.
We now consider the experiment with $L_{12}= C_{284}H_{190}F_{320}N_4S_{12}$ molecules reported in \cite{C3CP51500A}. The experimental setup is shown in Fig.~\ref{fig:experimental_setups}. The Fourier coefficients, defined in Eqs.~\eqref{t1} and \eqref{t3} for the transmission functions of the mechanical gratings, can be calculated analytically: $A_n=C_n=\frac{2l}{d}\sinc(\frac{l}{d}n)$. The velocity profile was approximated by a Gaussian centered around $v=85ms^{-1}$ with spread $\Delta v_{FWHM}=30ms^{-1}$ \cite{C3CP51500A}.

As an example, in Fig.~\ref{fig:interference_figures} we plot a comparison between the experimental interference pattern, the quantum mechanical fit and the CSL fit, for an arbitrarily chosen pair of parameters $\lambda, r_C$. We repeated the simulation for different pairs of parameters $\lambda, r_C$ as previously described. We obtain the parameter diagram shown in Fig.~\ref{CSL_parameter_diagram}.\\

\noindent {\bf Comparison of near and far field experiments}.
Fig.~\ref{CSL_parameter_diagram} shows the exclusion zone of the CSL parameters $\lambda$,$r_C$ for the far and near field experiments here considered. As we can see, they are similar: the near-field experiment sets a bound which is roughly two orders of magnitude stronger than the far-field experiment. This can be understood by the following argument.

Let us fix $r_C$ and focus our attention on the CSL model. The only remaining parameter is $\lambda$. We expect that deviations from standard quantum mechanics become important as $\lambda t$ increases. For the far-field experiment we have a typical flight time $t \approx 5 \text{ms} $ and molecular mass $m \approx 500 \text{amu}$. For the near-field experiment we have a typical flight time $t \approx 2 \text{ms} $  and  molecular mass $m \approx 10000 \text{amu}$.
Hence the ratio of bounds on $\lambda$ from the two experiments is approximately:
\begin{equation}
\frac{\lambda t |_{KDTL}}{\lambda t |_{far}}=\frac{(10000amu)^2 \;2ms}{ (500amu)^2 \;5ms} \approx 100.
\end{equation}
This rough estimate provides a simple explanation why the KDTL near-field experiment gives bounds which are 2 order of magnitude stronger than the bounds obtained from the far-field experiment.\\

\noindent {\bf Parameter bounds}.
The analysis here above refers to the CSL model, but easily applies to the other models discussed in this work. The bounds reported in Fig.~\ref{CSL_parameter_diagram} for the CSL model, refer also to dCSL an cCSL. The figure also shows the bounds coming from requiring that macroscopic objects are always well localized (see Sec.~\ref{sec_amp}). This bound puts the original value proposed by GRW right on the border of the exclusion zone (the shaded zone at the bottom). 

The dCSL bounds from interferometry change slightly if we consider very low temperatures or very high boosts. However, as already stressed before, the smallest modification of the quantum mechanical interference pattern is given by the dCSL model with infinite temperature and no boost, i.e. the CSL model. Hence, since we do not know the temperature and speed of the noise, the most conservative bounds for all dCSL models coincide with the CSL bounds. On the other hand, the bounds obtained by requiring that macroscopic objects are always well localized, become weaker for lower temperatures and higher boost of the noise, but this affects only very high values of $r_C$. On the other extreme, for very small $r_C$ values, the bounds for dCSL models with very low temperature may become invalid, as the approximations utilized begin to break down. See Figs.~\ref{limitsT} and \ref{limitsTB} for a quantitative analysis.

The cCSL bounds from interferometry experiments are valid for noises with a frequency cut-off $\Omega \gg 10^{13}Hz$. For comparison, bounds from X-ray experiments \cite{Curceanu}, refer to the cCSL model with a frequency cut-off  $\Omega \gg 10^{18}Hz$. For completeness, we have also shown the CSL bounds from LISA Pathfinder~\cite{carlesso2016experimental}. 

The fact that the CSL, cCSL and dCSL bounds in Fig.~\ref{CSL_parameter_diagram} coincide is due to the fact the time scale of dissipative and non-Markovian effects are longer than the experimental times. This result, shows that interferometric experiments provide bounds that are insensitive to dissipative or non-markavian extensions of the original models for very lage values of the parameters. Interferometric experiments can thus provide a test, not only for a specific model, but for a large class of collapse models, even  those not yet considered, such as a CSL model which is both dissipative and non-markovian. For a detailed analysis of the CSL, cCSL and dCSL bounds from non-interferometric experiments with cold-atoms see~\cite{bilardello2016bounds}.

The bounds for the GRW and dGRW models can be obtained from the bounds of the CSL and dCSL models, respectively, by changing the amplification factor $\Lambda$.

\begin{figure}
\includegraphics[width=0.75\textwidth]{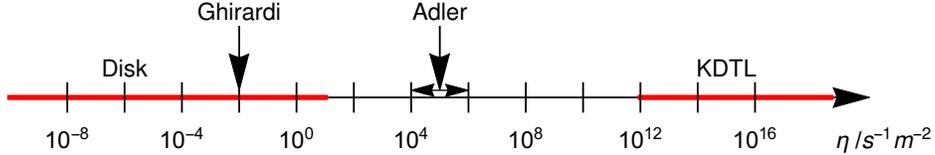}\centering\caption[QMUPL parameter diagram]{Bounds on the parameter $\eta$ of the QMUPL model from matter-wave interferometry (denoted
by KDTL) and from the request of suppression of macroscopic superpositions
(denoted by Disk). The excluded values are denoted by red lines. }
\label{fig_QMUPL}
\end{figure}

The bounds on the QMUPL model parameter $\eta$ are shown in Fig.
\ref{fig_QMUPL}. We can obtain some reference values for the parameter
$\eta$ in the following way: the QMUPL model can be obtained as the
limit of the GRW/CSL model~\cite{durr2011stochastic}, specifically, we have $\eta=\lambda/(2r_{C}^{2})$.
Using the values suggested in \cite{PhysRevD.34.470} we obtain
($\lambda=10^{-16}\text{s}^{-1}$, $r_{C}=10^{-7}\text{m}$): 
$\eta_{\text{\text{GRW}}}=  10^{-2}\text{s}^{-1}\text{m}^{-2}$.
We will refer to these value as the Ghirardi values.  In \cite{1751-8121-40-12-S03}
we have two different choices: $\lambda=10^{-8\pm2}\text{s}^{-1}$
($\lambda=10^{-6\pm2}\text{s}^{-1}$) and $r_{C}=10^{-7}\text{m}$
($r_{C}=10^{-6}\text{m}$). These give the following value: $
\eta_{\text{Adler}}=  10^{5\pm2}\text{s}^{-1}\text{m}^{-2}$.
We will refer to this value as the Adler value. From Fig.~\ref{fig_QMUPL} we see that the Ghirardi value is excluded by the requirement of macroscopic localization.

The KDTL bounds on the DP parameter $R_0$ fall below the regime of applicability of the DP model ($R_0\geq10^{-15}m$). In fact, the effective collapse rate of the DP model $\Lambda_{DP}=\frac{G m_0^2}{\hbar \sqrt{\pi} R_0} \frac{\lambda(R_0)}{\lambda}$ 
is very small above $10^{-15}m$, e.g. for $R_0=10^{-15}m$ we have $\Lambda_{DP} \approx 10^{-15}s^{-1} \frac{\lambda(R_0)}{\lambda}$, 
while for $R_0=10^{-7}m$ we have $\Lambda_{DP} \approx 10^{-23} s^{-1} \frac{\lambda(R_0)}{\lambda}$, which is orders of magnitude below the CSL bounds 
$\lambda \approx 10^{-3}s^{-1} (\lambda \approx 10^{-6} s^{-1}$) for  $r_C=10^{-15}m$ ($r_C=10^{-7}m$), respectively. On the other hand, the requirement  that macroscopic objects are always well localized provides very strong bounds. If we require that a single layered Graphene disk of radius $r=0.01\text{mm}$ is to be localized within $t=10\text{ms}$, as we have done for the CSL family of models, we can already exclude all values of $R_0$. However, even if we consider a larger value, for example $r=1\text{mm}$, the values $R_0=10^{-15}m$, $R_0=10^{-7}m$ proposed by  Di\'{o}si~\cite{Diosi1987377} and Ghirardi~\cite{ghirardi1990continuous}, respectively, are still excluded.

\clearpage
\section{Summary}\label{conclusions}
We have discussed the bounds on the parameters of the most well-known collapse models, arising from the most relevant matter-wave macromolecule experiments: the far field~\cite{Real} and the near field KDTL~\cite{C3CP51500A} experiments. We have derived the interference pattern using a density matrix formalism, where we have addressed the validity of the paraxial approximation, the seprability of the dynamics along the three axes and the amplification mechanism. 

One of the main results is that the bounds obtained for the standard CSL model are bounds also for more general, dissipative (dCSL) or non-markovian generalizations (cCSL). This is in contrast with other indirect experiments, which test only particular collapse models. 

In addition, we have seen that the localization requirement of a Graphene disk provides very stringent lower bounds on the collapse parameters. Specifically, the values proposed by Ghiradi, Rimini and Weber for the CSL model are right at the border, while the same requirement excludes entirely the DP model.

\section*{Acknowledgements}
The authors acknowledge financial support from the EU project NANOQUESTFIT, INFN and FRA 2016 (UNITS). We thank Prof. Markus Arndt for the data and the setup parameters of the experiment in \cite{MSCH} and for useful discussions. We also thank Prof. Hendrik Ulbricht for many enlightening discussions.

\clearpage
\bibliographystyle{unsrt}
\bibliography{diffraction.bib}

\end{document}